\documentclass[12pt]{article}
\usepackage[utf8]{inputenc}
\usepackage{color}
\usepackage{authblk}
\usepackage{latexsym}
\usepackage{epsfig,amssymb,euscript, mathrsfs,cite}
\usepackage{amsmath}
\usepackage{mathrsfs} 
\usepackage{enumerate}
\definecolor{MyDarkBlue}{rgb}{0.15,0.15,0.45}
\usepackage[linktocpage=true]{hyperref}
\hypersetup{
colorlinks=true,
citecolor=MyDarkBlue,
linkcolor=MyDarkBlue,
urlcolor=MyDarkBlue,
pdfauthor={},
pdftitle={},
pdfsubject={hep-th}
}
\newsavebox{\ns}
\newsavebox{\dbrane}
\newsavebox{\dbshort}

\def\be{\begin{equation}}
\def\ee{\end{equation}}
\def\bea{\begin{eqnarray}}
\def\eea{\end{eqnarray}}

\newcommand{\nn}{\nonumber}

\newcommand\cF{\mathcal{F}}
\newcommand\cA{\mathcal{A}}
\newcommand{\ma}{\mathrm{a}}

\newcommand{\mf}{\mathrm{f}}

\newcommand\R{\mathbb{R}}
\newcommand\Z{\mathbb{Z}}

\newcommand\C{\mathbb{C}}

\newcommand\diff{\mathrm{d}}

\newcommand{\dd}{\mathrm{d}}
\newcommand{\ii}{\mathrm{i}}

\newcommand{\ex}{\mathrm{e}}
\newcommand{\vol}{\mathrm{vol}}

\def\beq{\begin{equation}}
\def\eeq{\end{equation}}
\def\bea{\begin{align}} 
\def\eea{\end{align}}

\newlength{\sswidth}
\newcommand{\sla}[1]{
   \settowidth{\sswidth}{$#1$}
   \mbox{$\not{\hspace*{-0.15\sswidth}#1}$}}

\newcommand{\Xb}{X_1}
\newcommand{\Xs}{X_2}
\newcommand{\SUsign}{-}

\newcommand{\GG}{\mathscr{G}}
\newcommand{\met}{g}
\newcommand{\Met}{G}
\newcommand{\B}{Y}
\newcommand{\cC}{\mathcal{C}}
\newcommand{\cG}{\mathcal{G}}
\newcommand{\mexp}{\mathtt{g}}
\newcommand{\cutoff}{\delta}
\newcommand{\los}{\varepsilon}
\newcommand{\te}{\mathtt{e}}
\newcommand{\BJ}{\mathcal{J}}
\newcommand{\Jb}{\mathrm{J}}
\newcommand{\Ib}{\mathrm{I}}
\newcommand{\Jcur}{\mathscr{J}}
\newcommand{\mc}[1]{\mathcal{#1}}
\newcommand{\ph}[1]{\phantom{#1}}
\newcommand{\BK}{\mathcal{K}}
\newcommand{\Ex}{\mathrm{E}}
\newcommand{\Sx}{\mathscr{E}}
\newcommand{\aIfour}{a^I_1}
\newcommand{\aIfive}{a^I_2}
\newcommand{\mafour}{\ma_1}
\newcommand{\mafive}{\ma_2}
\newcommand{\Xseven}{X_3}
\newcommand{\Xeight}{X_4}
\newcommand{\mffive}{\mf_2}
\newcommand{\JRcur}{\mathbb{J}}

\numberwithin{equation}{section}       

\begin{document}

\begin{titlepage}

\begin{center}

\today

\vskip 2.3 cm 

\vskip 5mm

{\Large \bf Topological AdS/CFT}

\vskip 15mm

{Pietro Benetti Genolini${}^a$, Paul Richmond${}^b$ and James Sparks${}^a$}

\vspace{1cm}
\centerline{${}^a${\it Mathematical Institute, University of Oxford,}}
\centerline{{\it Andrew Wiles Building, Radcliffe Observatory Quarter,}}
\centerline{{\it Woodstock Road, Oxford, OX2 6GG, UK}}
\vspace{1cm}
\centerline{${}^b${\it INFN, sezione di Milano-Bicocca, I-20126 Milano, Italy}}

\end{center}

\bigskip
\begin{center}
{\bf {\sc Abstract}} 
\end{center}
We define a holographic dual to the Donaldson-Witten topological twist of $\mathcal{N}=2$ gauge theories on a Riemannian four-manifold. This is described by a class of asymptotically locally hyperbolic solutions to  $\mathcal{N}=4$ gauged supergravity in five dimensions, with the four-manifold as conformal boundary. Under AdS/CFT, minus the logarithm of the partition function of the gauge theory is identified with the holographically renormalized supergravity action. We show that the latter is independent of the metric on the boundary four-manifold, as required for a topological theory. Supersymmetric solutions in the bulk satisfy first order differential equations for a twisted $Sp(1)$ structure, which extends the quaternionic K\"ahler structure that exists on any Riemannian four-manifold boundary.  We comment on applications and extensions, including generalizations to other topological twists.
\end{titlepage}

\pagestyle{plain}
\setcounter{page}{1}
\newcounter{bean}
\baselineskip18pt
\tableofcontents


\section{Introduction and outline}\label{SecIntroduction}

The AdS/CFT correspondence is a conjectured duality relating certain quantum field theories (QFTs) to quantum gravity \cite{Maldacena:1997re}. This typically 
relates a strong coupling limit in field theory to semi-classical gravity, and quantitative comparisons between the two sides 
usually rely on additional symmetries, such as supersymmetry or integrability. Starting with the work of \cite{Pestun:2007rz}, 
recently localization techniques in supersymmetric gauge theories defined on rigid supersymmetric backgrounds have led to new exact 
computations. Moreover, the appropriate strong coupling limits have been successfully matched 
to semi-classical gravity calculations, in a variety of different set-ups.\footnote{A review of some of these results 
appears in \cite{Pestun:2016zxk}, although many more results have appeared since.} On the other hand, 
localization in QFT originated in \cite{Witten:1988ze}, where the topological twist was introduced to define 
a topological quantum field theory (TQFT). 
It is natural to then ask whether one can define and study holography 
in this topological setting. Indeed, what does gravity tell us about TQFT, and {\it vice versa}?
In this paper, we take some first steps in this direction.

\subsection{Background}\label{SecBackground}

In  \cite{Witten:1988ze}, Witten gave a physical construction of Donaldson invariants 
of four-manifolds \cite{Donaldson:1983wm, Donaldson:1990kn, DonKron} as certain correlation functions in a TQFT.  
This theory is 
 constructed by taking pure $\mathcal{N}=2$ Yang-Mills gauge theory and applying 
 a topological twist: identifying a background $SU(2)$ R-symmetry gauge field with the 
 right-handed spin connection results in a conserved scalar supercharge $\mathcal{Q}$, 
 on any oriented Riemannian four-manifold $(M_4,\met)$. 
  The path integral 
    localizes onto Yang-Mills instantons, 
and correlation functions of $\mathcal{Q}$-invariant operators localize to integrals of certain  forms over the instanton moduli space $\mathcal{M}$. 
These are precisely Donaldson's invariants of $M_4$. They are, under certain general conditions, 
independent of the choice of metric $\met$ on $M_4$, but in general depend on the diffeomorphism type of $M_4$.
In particular, Donaldson invariants can sometimes distinguish manifolds which are homeomorphic but not diffeomorphic. 
That this is possible is because 
the instanton equations are PDEs, which depend on the differentiable structure. From the TQFT point of view, independence of the choice 
of metric follows by showing that metric deformations lead to $\mathcal{Q}$-exact changes in the integrand of the path integral. 
For example, the stress-energy tensor is $\mathcal{Q}$-exact, implying that the partition function is invariant under arbitrary metric deformations, and hence (formally at least) is a diffeomorphism invariant. 

Donaldson-Witten theory is typically studied for pure $\mathcal{N}=2$ Yang-Mills, with gauge group $\GG=SU(2)$ or $\GG=SO(3)$. 
However, the topological twist may be applied to any $\mathcal{N}=2$ theory with matter, and also for any gauge group $\GG$. 
For example, $\GG=SU(N)$ Donaldson invariants were first studied in
\cite{Marino:1998bm}, with further mathematical work in \cite{kron}. In particular the latter reference contains 
some explicit large $N$ results for the partition function on certain four-manifolds.
The procedure of topological twisting may also be applied to theories with different amounts of supersymmetry, and in 
various dimensions. For example, the larger $SU(4)$ R-symmetry of four-dimensional $\mathcal{N}=4$ Yang-Mills 
leads to three  inequivalent twists \cite{Yamron:1988qc}. 
Viewing the $\mathcal{N}=4$ theory as an $\mathcal{N}=2$ theory coupled to an adjoint matter multiplet, 
applying the Donaldson-Witten twist leads to a TQFT that is referred to as the ``half-twisted'' $\mathcal{N}=4$ theory. 
This theory is relevant for the construction in the present paper.
The other two twists are the Vafa-Witten twist \cite{Vafa:1994tf}, and the twist studied by 
Kapustin-Witten in \cite{Kapustin:2006pk}, relevant for the 
Geometric Langlands programme.
Historically the development of Donaldson-like invariants
took a rather different direction after the introduction of 
Seiberg-Witten invariants in \cite{Witten:1994cg}. The former may be expressed (conjecturally) in terms of the latter, 
but Seiberg-Witten theory is  simpler and easier to compute with.

The Donaldson-Witten twist of $\mathcal{N}=2$ gauge theories can be understood 
as a special case of rigid supersymmetry. Soon after Witten's paper, Karlhede-Ro\v{c}ek 
interpreted the construction as coupling the gauge theory to a background ({\it i.e}.\ non-dynamical) $\mathcal{N}=2$ 
conformal gravity \cite{Karlhede:1988ax}. The background $SU(2)$ R-symmetry gauge field is 
part of this gravity multiplet, and is embedded into the spin connection in such a way that 
the Killing spinor equations of the theory  admit a constant solution, leading to the conserved scalar supercharge $\mathcal{Q}$. 
There is also an auxiliary scalar field turned on in this background gravity multiplet, proportional to the Ricci scalar 
curvature of $(M_4,\met)$. Motivated by the work of Pestun in \cite{Pestun:2007rz}, 
the last few years have seen considerable interest in defining rigid supersymmetry more generally on Riemannian manifolds. 
Unlike the topological twist, this generally requires the background $d$-manifold $(M_d,\met)$ to possess some additional geometric structure, 
and correlation functions of $\mathcal{Q}$-invariant observables then usually depend on this structure.
For example, one can couple four-dimensional 
$\mathcal{N}=1$ theories with a $U(1)$ R-symmetry to a background new minimal supergravity.
Geometrically this construction requires
 $(M_4,\met)$ to be  a Hermitian four-manifold, with an integrable complex structure
\cite{Dumitrescu:2012ha, Klare:2012gn}. Generalizing \cite{Karlhede:1988ax}, 
similarly $\mathcal{N}=2$ theories may be coupled to a background $\mathcal{N}=2$ conformal 
supergravity \cite{Klare:2013dka}. Generically this requires 
the existence of a conformal Killing vector on $(M_4,\met)$, but the 
topological twist arises as a degenerate special case, in which $(M_4,\met)$ is arbitrary. 

An interesting application of these constructions is to the AdS/CFT correspondence. Here strong coupling 
(typically large rank $N$) gauge theory computations are related to semi-classical gravity. 
The general idea is as follows. Rigid supersymmetry generically equips the background manifold $(M_d,\met)$, on which the 
gauge theory is defined, 
with certain additional geometric structure, such as the integrable complex structure mentioned 
for four-dimensional $\mathcal{N}=1$ theories above. In the gravitational 
dual description one seeks solutions to an appropriate supergravity theory in $d+1$ dimensions, 
where $(M_d,\met)$ arises as a conformal boundary. That is, the $(d+1)$-dimensional 
metric is asymptotically locally hyperbolic, approximated by 
$\frac{\diff z^2}{z^2} + \frac{1}{z^2}g$ to leading order in $z$ near the conformal boundary at $z=0$.  
A saddle point approximation to quantum gravity in this bulk then identifies
\beq\label{saddle}
Z[M_d] \ = \ \sum \ex^{-S[\B_{d+1}]}~.
\eeq
Here $Z[M_d]$ denotes the partition function of the gauge theory defined on $M_d$, while 
$S[\B_{d+1}]$ is the holographically renormalized supergravity action, evaluated 
on an asymptotically locally hyperbolic solution to the equations of motion of the $(d+1)$-dimensional 
theory. The manifold $M_d=\partial \B_{d+1}$ is the conformal boundary, with  the boundary conditions 
for supergravity fields on $\B_{d+1}$ fixed by the rigid background structure of $M_d$. 

The general AdS/CFT relation (\ref{saddle}) is somewhat schematic, and both sides 
must be interpreted appropriately. For example, in order to make sense of the left hand side 
for topologically twisted four-dimensional $\mathcal{N}=2$ SCFTs it can be refined, as discussed in section \ref{SecTopAdSCFT}. 
On the other hand, the sum on the right hand side of (\ref{saddle}) is not well understood. One should certainly include 
all saddle point solutions on smooth manifolds $\B_{d+1}$. However, the existence of such a filling immediately implies 
that $M_d$  has trivial class in the oriented bordism group $\Omega_d^{SO}$, in general constraining the choice of $M_d$.\footnote{For example, in the case of interest in this paper $d=4$, and $\Omega_4^{SO}\cong \Z$ 
with the map to the integers being given by the signature $\sigma(M_4)=b_2^+(M_4)-b_2^-(M_4) 
= \frac{1}{3}\int_{M_4}p_1(M_4)$, where $p_1$ denotes the first Pontryagin class. 
A generator of $\Omega_4^{SO}\cong \Z$ is 
 the complex projective plane.} 
That said, various explicit examples (see, for example, \cite{Alday:2012au, Alday:2015jsa, Banerjee:2009af}) suggest that requiring $\B_{d+1}$ 
  to be smooth is in any case too strong: one should allow for certain types of singular
  fillings of $(M_d,\met)$, 
   and indeed these may even be the dominant contribution 
  in (\ref{saddle}) (especially for non-trivial topologies of $M_d$).
   There are some clear constraints, although no 
  general prescription.\footnote{One might also speculate  that the dominant contribution may come from complex saddle points; that is, from  complex-valued metrics -- see, for example, \cite{Maloney:2007ud}. In this paper we focus on real solutions.} The supergravity action $S$ typically scales with a positive power of $N$, and 
  in the $N\rightarrow\infty$ limit only the solution of least action contributes to (\ref{saddle}) 
  at leading order, with contributions from other solutions being exponentially suppressed.

\subsection{Outline}\label{SecOutline}

In this paper we construct a holographic dual to the Donaldson-Witten twist of 
four-dimensional $\mathcal{N}=2$ gauge theories. As already mentioned, this 
twist may be interpreted as coupling the theory to a particular background $\mathcal{N}=2$ 
conformal gravity multiplet. On the other hand, four-dimensional $\mathcal{N}=2$ conformal gravity arises 
on the conformal boundary of asymptotically locally hyperbolic solutions to 
the Romans \cite{Romans:1985ps} $\mathcal{N}=4^+$ gauged supergravity in five dimensions 
\cite{Ohl:2010au}. The real Euclidean signature version of this theory described in section \ref{SecRomansSUGRA} 
has, in addition to the bulk metric $\Met_{\mu\nu}$, an $SU(2)$ R-symmetry gauge field $\cA^I_\mu$ ($I=1,2,3$), a one-form $\cC$, and a scalar field $X$. (In general 
there is also a doublet of $B$-fields, but this is zero for the topological  twist boundary condition, and moreover 
may be consistently set to zero in the Romans theory.)

The main property of a topological field theory is that appropriate correlation functions, 
including the partition function, are independent of any choice of metric.  
Assuming one is given an appropriate solution to the Romans theory with $(M_4,g)$ 
as conformal boundary, we therefore expect the holographically renormalized action to be 
independent of $g$. Here  one can mimic the field theory argument in \cite{Witten:1988ze}, and 
attempt to show that arbitrary deformations $g_{ij}\rightarrow g_{ij}+\delta g_{ij}$ leave this action invariant. We have the 
general holographic Ward identity formula
\beq\label{holWard}
\delta S \ = \ \int_{M_4} \diff^4 x \sqrt{\det g} \left(\tfrac{1}{2}T_{ij}\, \delta g^{ij} + \Jcur_I^i\, \delta A_i^I + \Xi\, \delta \Xb\right)~.
\eeq
Here $S$ is the renormalized supergravity action of the Euclidean Romans theory, defined in section \ref{SecRomansSUGRA}, while $(g_{ij}, A^I_i, \Xb)$ are 
the non-zero background fields in the $\mathcal{N}=2$ conformal gravity multiplet 
for the topological twist. Equivalently, these arise as boundary values of the Romans fields: 
in particular $A^I_i$ is simply the restriction of the bulk $SU(2)$ R-symmetry gauge field to the boundary at $z=0$, 
while $\Xb=\lim_{z\rightarrow 0} (X-1)/z^2\log z$.
For the topological twist these quantities are all fixed by the choice of metric $g_{ij}$: 
$A^I_i$  is fixed to be the right-handed spin connection, while $\Xb=-R/12$, where 
$R=R(g)$ is the Ricci scalar for $g$. Thus the variations of these fields appearing in (\ref{holWard}) 
are all determined by the metric variation $\delta g_{ij}$.
On the other hand, $T_{ij}$, $\Jcur_I^i$ and $\Xi$ are respectively the 
holographic vacuum expectation values (VEVs) of the operators for which these boundary fields are the sources. 
In particular $T_{ij}$ is the holographic stress-energy tensor. As is well-known, 
the expansion of the equations of motion near $z=0$ does not fix these VEVs in terms of boundary data on $M_4$, but rather they
are only determined by regularity of the solution in the interior. Determining these quantities for fixed boundary data
is thus an extremely non-linear problem. What allows progress 
in this case is supersymmetry: the partition function should be described by a supersymmetric 
solution to the Romans theory.\footnote{If the dominant saddle point in (\ref{saddle}) were non-supersymmetric, 
this would presumably be interpreted as spontaneous breaking of supersymmetry in the dual TQFT. 
This is certainly not expected in the case at hand, but would be interesting to investigate further.}
By similarly solving the Killing spinor equations in a Fefferman-Graham-like expansion, 
we are able to compute these VEVs for a general supersymmetric solution. This 
still leaves certain unknown data, ultimately determined by regularity in the interior,
 but remarkably these constraints are sufficient to prove that (\ref{holWard}) 
 is indeed zero, for arbitrary $\delta g_{ij}$! More precisely, we show that the integrand on the right hand side is
 a total derivative, and its integral is then 
 zero provided $M_4$ is closed, without boundary. The computation, although 
 in principle straightforward, is not entirely trivial, and along the way we require 
some 
 interesting identities that are specific to Riemannian four-manifolds (notably  the quadratic curvature 
 identity of Berger \cite{Berger}).
This is the main result of the paper, but it immediately raises a number of  interesting questions. 
We postpone our discussion of these until later in the paper, notably at the end of section \ref{SecVary}, 
and in sections \ref{SecGeometry} and \ref{SecDiscussion}.
 
The outline of the paper is as follows.  In section \ref{SecRomansSUGRA} we define the relevant 
five-dimensional Euclidean $\mathcal{N}=4^+$ gauged supergravity theory, and holographically 
renormalize its action $S$. In section \ref{SecSUSY} we show that on the conformal boundary 
of an asymptotically locally hyperbolic solution to this theory one obtains the supersymmetry 
equations \cite{Klare:2013dka} of Euclidean $\mathcal{N}=2$ conformal supergravity, which 
admits \cite{Karlhede:1988ax} the topological twist  as a solution. We then expand 
the bulk supersymmetry equations in a Fefferman-Graham-like expansion. 
Section \ref{SecVary} contains the main proof that $\delta S/\delta g_{ij}=0$, while in 
section \ref{SecGeometry} we reformulate the supersymmetry equations 
in terms of a first order differential system for a twisted $Sp(1)$ structure. On the 
conformal boundary this induces the canonical quaternionic K\"ahler structrure that 
exists on any oriented Riemannian four-manifold. This paper raises a number of 
interesting questions, prompting further computations, and the results 
may potentially be extended and generalized in a number of different directions. 
We comment on some of these issues in section \ref{SecDiscussion}.

\section{Holographic supergravity theory}\label{SecRomansSUGRA}

We begin in section \ref{SecRomans} by defining a real Euclidean section of $\mathcal{N}=4^+$ gauged supergravity 
in five dimensions. A Fefferman-Graham expansion of asymptotically locally hyperbolic solutions 
to this theory is constructed in section \ref{SecFG}, for arbitrary conformal boundary four-manifold $(M_4,\met$). 
Using this, in section \ref{SecHoloRenormalization} we holographically renormalize the action.

\subsection{Euclidean Romans \texorpdfstring{$\mathcal{N}=4^+$}{N=4+} theory}\label{SecRomans}

The Lorentzian signature 
Romans $\mathcal{N}=4^+$ theory \cite{Romans:1985ps} is a five-dimensional $SU(2)\times U(1)$ gauged supergravity which admits a supersymmetric AdS$_5$ vacuum. It is a consistent truncation of both Type IIB supergravity on $S^5$ \cite{Lu:1999bw}, and also eleven-dimensional supergravity on an appropriate class of six-manifolds ${N}_6$ \cite{Gauntlett:2007sm}.  The bosonic sector comprises the metric $\Met_{\mu\nu}$, a dilaton $\phi$, an $SU(2)_R$ Yang-Mills gauge field $\cA^I_\mu$ ($I=1,2,3$), a $U(1)_R$ gauge field $\cA_\mu$, and two real anti-symmetric tensors $B^\alpha_{\mu\nu}$, $\alpha=4,5$, which transform as a charged doublet under $U(1)_R\cong SO(2)_R$. 
It is convenient to introduce the scalar field $X\equiv \ex^{-\frac{1}{\sqrt6}\, \phi}$ and the complex combinations $\mathcal{B}^\pm \equiv B^4 \pm \ii B^5$.  The associated field strengths are  $\cF = \dd \cA$, $\cF^I = \dd \cA^I \SUsign \frac{1}{2} \epsilon^{I}_{\ JK} \cA^J \wedge \cA^K$,  and $H^\pm = \dd \mathcal{B}^\pm \mp \ii  \cA \wedge \mathcal{B}^\pm$. We have set the gauged supergravity gauge coupling to $1$.\footnote{In addition we have rescaled the $SU(2)_R$ gauge field and the anti-symmetric tensors by a factor of $1/\sqrt{2}$, compared to 
\cite{Lu:1999bw}.}

The bosonic action and equations of motion in Lorentzian signature appear in \cite{Lu:1999bw}.  However, as we are interested in holographic duals to TQFTs defined on Riemannian four-manifolds, we require the Euclidean signature version of this theory. The Wick rotation 
in particular introduces a factor of $\ii$ into the Chern-Simons couplings, leading to the Euclidean action
\begin{align}
	I \ = & \  - \frac{1}{2\kappa_5^2} \int \ \Big[ R \, {*1} - 3 X^{-2} \dd X \wedge *\dd X + 4  ( X^2 + 2 X^{-1} ) \, {*1} - \tfrac{1}{2} X^4 \, \cF \wedge *\cF  \label{IEuclid}\\
	&\ - \tfrac{1}{4} X^{-2} \, ( \cF^I \wedge * \cF^I + {\mathcal{B}^-} \wedge * \mathcal{B}^+ )+ \tfrac{1}{8} {\mathcal{B}^-} \wedge H^+ - \tfrac{1}{8} \mathcal{B}^+ \wedge {H}^- - \tfrac{\ii}{4} \cF^I \wedge \cF^I \wedge \cA \Big] \nn.
\end{align}
Here $R=R(G)$ denotes the Ricci scalar of the metric $G_{\mu\nu}$, and $*$ is the Hodge duality operator acting on forms. 
The associated equations of motion are:\footnote{Equation \eqref{AIeom} incorporates a correction to the Lorentzian equation, in line with \cite{Gauntlett:2007sm}.}
\begin{align}
	\dd ( X^{-1}\, {*\, \dd X})   \  = & \ \tfrac{1}{3} X^4 \, \cF \wedge *\cF - \tfrac{1}{12} X^{-2} \, (\cF^I \wedge *\cF^I + {\mathcal{B}^-} \wedge *\mathcal{B}^+)\nn\\  &  - \tfrac{4}{3}  (X^2 - X^{-1})\, *1~, \label{Xeom} \\[5pt]
	\dd ( X^{-2} * \cF^I ) \ =& \  \   \epsilon^{I}_{\ JK} X^{-2} * \cF^J \wedge \cA^K - \ii \cF^I \wedge \cF~, \label{AIeom} \\[5pt]
	\dd ( X^4 * \cF ) \ =& \ - \tfrac{\ii}{4} \cF^I \wedge \cF^I - \tfrac{\ii}{4} {\mathcal{B}^-} \wedge \mathcal{B}^+~, \label{Aeom} \\[5pt]	
	H^\pm \ =& \ \pm X^{-2} * \mathcal{B}^\pm \label{Beom}~, \\[5pt]
	%
	%
	%
	%
	%
		R_{\mu\nu}  \ = & \  3  X^{-2} \partial_\mu X \partial_\nu X - \tfrac{4}{3} ( X^2 + 2 X^{-1} ) G_{\mu\nu}+ \tfrac{1}{2} X^4 \big( \cF_\mu{}^\rho \cF_{\nu\rho} - \tfrac{1}{6} G_{\mu\nu} \cF^2 \big)\ \ \ \ \ \ \ \ \nn \\
	& + \tfrac{1}{4} X^{-2} \big( \cF^I_\mu{}^\rho \cF^I_{\nu\rho} - \tfrac{1}{6} G_{\mu\nu} ( \cF^I )^2 + {\mathcal{B}^-}_{(\mu}{}^\rho \mathcal{B}^+_{\nu)\rho} - \tfrac{1}{6} G_{\mu\nu} {\mathcal{B}^-}_{\rho\sigma} \mathcal{B}^{+ \rho\sigma} \big) \label{geom}\, .
\end{align}
Here $\cF^2\equiv \cF_{\mu\nu}\cF^{\mu\nu}$, $( \cF^I )^2\equiv \sum_{I=1}^3\cF^I_{\mu\nu}\cF^{I\mu\nu}$.
In general  equations (\ref{Xeom})--(\ref{geom}) are complex, and solutions will likewise be complex. However, note that 
setting $\ii\cA \equiv \cC$ effectively removes all factors of $\ii$. We may then consistently define a real section of this Euclidean 
theory in which all fields, and in particular $\cC$ and  $\mathcal{B}^\pm=B^4 \pm \ii B^5$, are real. We henceforth impose these
reality conditions. 
Although globally $\cA$ is a $U(1)_R$ gauge field in the original Lorentzian theory, after the above Wick rotation 
the real field $\cC=\ii\cA$ effectively becomes an $SO(1,1)_R$ gauge field. We may then think of $\cC$ as a 
global one-form, but for which the theory has a symmetry $\cC\rightarrow \cC - \diff \lambda$, for 
any global function $\lambda$. We denote the corresponding field strength as $\cG\equiv \diff \cC = \ii \cF$.
 
In the Lorentzian theory the fermionic sector contains four gravitini and four dilatini, which together with the spinor parameters $\epsilon$ all  transform in the fundamental $\mathbf{4}$ representation of 
the $Sp(2)_R$ global R-symmetry  group. The $SU(2)\times U(1)\subset Sp(2)$ gauge symmetry arises as a gauged subgroup. 
Since $Sp(2)\cong {Spin}(5)$ it is natural to introduce the associated Clifford algebra $\mathrm{Cliff}(5,0)$, with 
generators $\Gamma_A$, $A=1,\ldots,5$, satisfying $\{\Gamma_A,\Gamma_B\}=2\delta_{AB}$.  
We then decompose $I,J,K=1,2,3$, transforming in the $\mathbf{3}$ of $SU(2)$, and $\alpha, \beta = 4, 5$ in the $\mathbf{2}$ of $U(1)$. 
In Euclidean signature the conditions for preserving supersymmetry are then the vanishing of the following supersymmetry variations 
of the gravitini and dilatini, respectively: 
\begin{align}
0\ =& \ D_\mu \epsilon + \tfrac{\ii}{3} \gamma_\mu \Big( X + \tfrac{1}{2} X^{-2} \Big)  \Gamma_{45}  \epsilon \nn \\
	& \ + \tfrac{\ii}{24} ( \gamma_\mu{}^{\nu\rho} - 4 \delta_\mu^\nu \gamma^\rho ) \left( X^{-1} \big(\cF_{\nu\rho}^I\Gamma_I + B^\alpha_{\nu\rho} \Gamma_\alpha  \big) + X^2 \mathcal{F}_{\nu\rho} \right) \epsilon~, \label{BulkGravitino} \\
0\  =& \ \tfrac{\sqrt{3}}{2} \ii \gamma^\mu  X^{-1} \partial_\mu X \epsilon + \tfrac{1}{\sqrt{3}} \Big( X - X^{-2} \Big)  \Gamma_{45}  \epsilon \nn \\
	& \ + \tfrac{1}{8\sqrt{3}} \gamma^{\mu\nu} \Big( X^{-1} \big( \cF_{\mu\nu}^I  \Gamma_I   + B^\alpha_{\mu\nu} \Gamma_\alpha  \big) - {2} X^2 \mathcal{F}_{\mu\nu} \Big) \epsilon \label{BulkDilatino} \, ,
\end{align}
where the covariant derivative is
\beq
D_{\mu}\epsilon \ \equiv \ \nabla_{\mu}\epsilon + \tfrac{1}{2} \cA_{\mu}\Gamma_{45}\epsilon + \tfrac{1}{{2}} \cA^I_{\mu}\Gamma_{I45}\epsilon~.
\eeq
Here $\gamma_\mu$, $\mu=1,\ldots,5$, are generators of the Euclidean spacetime Clifford algebra, satisfying $\{\gamma_\mu,\gamma_\nu\}=2\Met_{\mu\nu}$, where recall $\Met_{\mu\nu}$ is the metric. Given the gauging it is natural to introduce the following choice of generators:
\beq
\Gamma_I \ = \ \sigma_3\otimes \sigma_I~, \quad I=1,2,3\, , \qquad \Gamma_4 \ = \ \sigma_1\otimes 1_2~, \qquad \Gamma_5 \ = \ 
\sigma_2 \otimes 1_2~,
\eeq
where $\sigma_I$ are the Pauli matrices, and $1_2$ denotes the $2\times 2$ identity matrix. 
In particular notice that $\Gamma_{45}=\ii \sigma_3\otimes 1_2$ squares to $-1_4$, and we may 
write 
\beq
\epsilon \ = \ \begin{pmatrix} \epsilon^+ \\ \epsilon^- \end{pmatrix}~,\label{pmeigen}
\eeq
where the spinor doublets $\epsilon^\pm$ denote projections onto the $\pm \ii$ eigenspaces 
of $\Gamma_{45}$, respectively. 
One then has
\beq
\Gamma_I \epsilon \ = \ \begin{pmatrix} \sigma_I \epsilon^+ \\ -\sigma_I \epsilon^-\end{pmatrix}~, \qquad 
B_{\mu\nu}^\alpha \Gamma_\alpha\epsilon \ =  \ \begin{pmatrix} \mathcal{B}_{\mu\nu}^-\epsilon^- \\ \mathcal{B}_{\mu\nu}^+ \epsilon^+\end{pmatrix}~.
\eeq
We next introduce the charge conjuguation matrix $\mathscr{C}$ for the Euclidean spacetime Clifford algebra. By 
definition $\gamma_\mu^*=\mathscr{C}^{-1}\gamma_\mu \mathscr{C}$, and one may choose Hermitian 
generators $\gamma_\mu^\dagger=\gamma_\mu$ together with the conditions 
$\mathscr{C}=\mathscr{C}^*=-\mathscr{C}^T$, $\mathscr{C}^2=-1$. We may then define the following 
charge conjugate spinor in Euclidean signature
\beq\label{Bulkcc}
\epsilon^c \ \equiv \ \left(\sigma_3\otimes \ii \sigma_2\right) \mathscr{C}\epsilon^*~.
\eeq
It is straightforward to check that $(\epsilon^c)^c=\epsilon$. Moreover, provided 
$\cC=\ii \cA$ and $\mathcal{B}^\pm$ (and all other bosonic fields) are real, then 
one can show that 
$\epsilon$ satisfies the 
gravitini and dilatini equations (\ref{BulkGravitino}), (\ref{BulkDilatino}) if and only if its
charge conjugate $\epsilon^c$ satisfies the same equations. Given this property, 
we may consistently impose the symplectic Majorana condition $\epsilon^c=\epsilon$. 
We will be interested in solutions that satisfy these reality conditions. 

\subsection{Fefferman-Graham expansion}\label{SecFG}

In this section we determine the Fefferman-Graham expansion \cite{Fefferman:2007rka} of asymptotically locally hyperbolic solutions to this Euclidean Romans theory. This 
is the general solution to the bosonic equations of motions (\ref{Xeom})--(\ref{geom}), expressed as a perturbative expansion in a radial coordinate near the conformal boundary. 

We take the form of the metric to be \cite{Fefferman:2007rka}
\begin{align}
	G_{\mu\nu}\diff x^\mu \diff x^\nu \ = \ \frac{1}{z^2} \dd z^2 + \frac{1}{z^2} \mexp_{ij} \dd x^i\dd x^j \ = \ \frac{1}{z^2} \dd z^2 + h_{ij} \dd x^i\dd x^j~. \label{FGmetric}
\end{align}
where the AdS radius $\ell=1$, and in turn  we have the expansion
\begin{align}
	\mexp_{ij} \ = \ \mexp_{ij}^{0}+z^2 \mexp_{ij}^{2}+z^4 \big( \mexp_{ij}^{4} + h_{ij}^{0} ( \log z )^2 + h_{ij}^{1} \log z \big) + o(z^4)~. \label{metricexp}
\end{align}
Here $\mexp_{ij}^{0}=\met_{ij}$ is the boundary metric induced on the conformal boundary $M_4$ at $z=0$. 

It is convenient to introduce the inner product $\langle \alpha , \beta \rangle$ between two $p$-forms $\alpha$, $\beta$ via
\begin{align}
	\alpha \wedge * \beta \ = \ \frac{1}{p!} \alpha_{\mu_1 \cdots \mu_p} \beta^{\mu_1 \cdots \mu_p}\,  \vol \ = \ \ \frac{1}{p!}\langle \alpha , \beta \rangle\,  \vol~,
\end{align}
where $\vol$ denotes the volume form, with associated Hodge duality operator $*$. 
The volume form for the five-dimensional bulk metric (\ref{FGmetric}) is 
\begin{align}
	\vol_5 \ =& \ \frac{1}{z^5}\dd z\wedge \vol_\mexp \ = \ \frac{1}{z^5}\dd z\wedge \sqrt{\det \mexp} \, \dd x^1\wedge \cdots \wedge \dd x^4~.\label{volform}
\end{align}
The determinant may then be expanded in a series in $z$, around that for $\mexp^0$, as follows
\begin{align}
\sqrt{\det \mexp} \ =  \sqrt{\det \mexp^0} \, \Big[ & 1 + \tfrac{z^2}{2} t^{(2)} + \tfrac{z^4}{2} \big( t^{(4)} - \tfrac{1}{2} t^{(2,2)} + \tfrac{1}{4} ( t^{(2)} )^2 \nn \\ & + u^{(0)} (\log z)^2 + u^{(1)} \log z \big) \Big] + o(z^4) \, .
\end{align}
Here we have denoted $t^{(n)} \equiv \mathrm{Tr} \left[ (\mexp^0)^{-1} \mexp^n \right]$, $u^{(n)}\equiv \mathrm{Tr} \left[ (\mexp^0)^{-1} h^n \right]$ and $t^{(2,2)} \equiv \mathrm{Tr} \left[ (\mexp^0)^{-1} \mexp^2 \right]^2$. 

The remaining bosonic fields are likewise expanded as follows:
\begin{align}
	X \ =& \ 1 + z^2 \left( \Xb \log z + \Xs \right) + z^4 (\Xseven \log z + \Xeight ) + o(z^4)~, \label{Xexp} \\[5pt]
	\cA^I \ =& \ A^I + z^2 (\aIfour \log z + \aIfive)  + o(z^2)~, \label{aIexp} \\[5pt]
	\cA \ =& \ \ma + z^2 (\mafour \log z + \mafive) + o(z^2)~, \label{aexp} \\[5pt]
	\mathcal{B}^\pm \ =& \ \frac{1}{z}b^\pm + \dd z \wedge {b}^\pm_1 + z(b_2^\pm \log z + b^\pm_3)  + o(z) \label{Bexp} ~,
\end{align}
\emph{A priori} there are additional terms that appear in these expansions. However, these may either be gauged away, or
turn out to be  set to zero by the equations of motion, and we have thus removed them in order to streamline the presentation.
 
We now substitute the above expansions into the equations of motion \eqref{Xeom}--\eqref{geom} and solve them order by order in the radial coordinate $z$ in terms of the boundary data $\mexp^0 = g, \Xb, A^I, \ma$ and $b^\pm$. This will leave a number of terms undetermined. For the Einstein equation (\ref{geom}) we will need the Ricci tensor of the metric \eqref{FGmetric}:
\begin{align}
	R_{zz} =&- \frac{4}{z^2}-\frac{1}{2}\Big(\mathrm{Tr}\left[\mexp^{-1}\partial^2_z\mexp\right]-\tfrac{1}{z}\mathrm{Tr}\left[\mexp^{-1}\partial_z\mexp\right]-\tfrac{1}{2}\mathrm{Tr}\left[\mexp^{-1}\partial_z\mexp\right]^2\Big)~,\label{Rzz}\\[5pt]
	R_{ij} =& -\frac{4}{z^2}\mexp_{ij}-\Big(\tfrac{1}{2}\partial^2_z\mexp-\tfrac{3}{2z}\partial_z\mexp-\tfrac{1}{2}(\partial_z\mexp)\mexp^{-1}(\partial_z\mexp)+\tfrac{1}{4}(\partial_z\mexp)\mathrm{Tr}\left[\mexp^{-1}\partial_z\mexp\right]\nn\\
	&\hspace{2.8cm} - R(\mexp)-\tfrac{1}{2z}\mexp\mathrm{Tr}\left[\mexp^{-1}\partial_z\mexp\right]\Big)_{ij}~,\label{Rij}\\[5pt]
	R_{zi} =& - \frac{1}{2}(\mexp^{-1})^{jk}\Big(\nabla_i\mexp_{jk,z}-\nabla_k\mexp_{ij,z}\Big)~.\label{Rzi}
\end{align}
Here $\nabla$ is the covariant derivative for $\mexp$, and we have corrected the sign of $R(\mexp)_{ij}$ and the right hand side of \eqref{Rzi} compared to \cite{Taylor:2000xw}. 

Examining first the equation (\ref{Beom}) gives at leading order
\beq
*_{\mexp^0} b^\pm \ = \ \mp  b^\pm \, , \label{b0Duality}
\eeq
so that the boundary $B$-fields $b^+,b^-$ are required to be anti-self-dual and self-dual, respectively. 
At subleading orders one finds
\begin{align}
{b}^\pm_1  \ =& \ \mp *_{\mexp^0} \left( \dd b^\pm \mp \ii \ma\wedge b^\pm \right) \, ,  \qquad 
	*_{\mexp^0}\, b^\pm_2 \ = \ \ \pm  ( b^\pm_2 - 2 \Xb \, b^\pm) \, .
	\end{align}
In particular notice that the first equation fixes $b_1^\pm$ in terms of boundary data, while the second equation 
determines only the anti-self-dual/self-dual parts of $b_2^\pm$, respectively. An 
equation may also be derived for $b_3^\pm$, although we will not need this in what follows.

Next the gauge field equations (\ref{AIeom}), (\ref{Aeom}) determine
\begin{align}
	\mafour \ =& \ - \tfrac{1}{2} \, *_{\mexp^0}  \dd *_{\mexp^0} \mathrm{f} + \tfrac{\ii}{8} *_{\mexp^0} \left(b^-\wedge {b}^+_1 + b^+\wedge b_1^-\right) \, ,\nn  \\
	\aIfour\ =& \ - \tfrac{1}{2} *_{\mexp^0} \mathcal{D} *_{\mexp^0} F^I \, , \label{a4solns}
	\end{align}
in terms of boundary data, 
where the curvatures are $\mathrm{f} \equiv \dd \ma$, $F^I \equiv \dd A^I - \frac{1}{{2}} \epsilon^{I}_{\ JK} A^J \wedge A^K$, and we have introduced a gauge covariant derivative with respect to the boundary $SU(2)$ field: $\mathcal{D} \alpha^I \equiv \dd \alpha^I \SUsign \epsilon^{I}_{\ JK} A^J\wedge \alpha^K$.
In addition we have the constraints
\begin{align}
	\dd *_{\mexp^0} \mafive \ =& \ - \tfrac{\ii}{8}F^I\wedge F^I - \tfrac{\ii}{8}\left(b^-\wedge b^+_3 + b^+\wedge b^-_3\right) \,~,  \qquad \quad \label{divJR}
	\mathcal{D} *_{\mexp^0}\aIfive \ = \ 0 ~,
\end{align}
which leave $\mafive$ and $\aIfive$ partially undetermined.

Turning next to the scalar equation of motion (\ref{Xeom}) we find
\begin{align}
	4\Xseven \, =& \, -\nabla^2 \Xb - 2 \left(t^{(2)}\Xb - 2 \Xb^2 \right) - \tfrac{1}{24}\left( \langle b^+, b^-_2 \rangle_{\mexp^0} + \langle b^-, b^+_2 \rangle_{\mexp^0} \right)~, \label{nablaXb} \\
 4\Xeight	\, =& \, -\nabla^2 \Xs - \left( t^{(2)}\Xb + 2t^{(2)}\Xs - \Xb^2 - 4\Xb \Xs + 4\Xseven\right)  - \tfrac{1}{24}\langle F^I, F^I \rangle_{\mexp^0}+ \tfrac{1}{6}\langle \mf, \mf \rangle_{\mexp^0}  \nn \\
	&- \tfrac{1}{12}\langle {b}^+_1, {b}^-_1\rangle_{\mexp^0} + \tfrac{1}{12}\langle b^-, \mexp^2\circ b^+\rangle_{\mexp^0} - \tfrac{1}{24}\left( \langle b^+, b^-_3 \rangle_{\mexp^0} + \langle b^-, b^+_3 \rangle_{\mexp^0}\right) \, . \label{nablaXs}
\end{align}
We regard these as determining $\Xseven$, $\Xeight$ in terms of $\Xb$ (a boundary field), and $\Xs$ (which is undetermined by the
equations of motion), together with the other fields in the expansion. 
In the second equation we have used the definition
\begin{align}
	(\mexp^2\circ\alpha)_{i_1\cdots i_p} \ \equiv \ (\mexp^2)_{[i_1}{}^{j} \alpha_{|j|i_2\cdots i_p]}~,
\end{align}
where $\alpha$ is a $p$-form on $M_4$.
Here indices are always raised with $\mexp^0$, so $(\mexp^2)_{i}{}^{j}\equiv (\mexp^2)_{ik}(\mexp^0)^{kj}$.

Finally, we introduce the matter modified boundary Ricci tensor 
\begin{align}
	\mathscr{R}_{ij} \ = \ \mathscr{R}_{ij}(\mexp^0) \ \equiv \ R_{ij} ( \mexp^0 ) - \tfrac{1}{4} ( b^+ )_{(i}{}^k ( b^- )_{j)k} ~.\label{BRicci}
\end{align}
Notice the scalar curvature is $\mathscr{R}(\mexp^0) = R(\mexp^0)$, due to the opposite duality properties 
(\ref{b0Duality}) of $b^\pm$.
From the $ij$ component of the Einstein equation \eqref{geom}, using \eqref{Rij} gives
\begin{align}
	\mexp^2_{ij} \ =& \ - \tfrac{1}{2} \big( \mathscr{R}_{ij} - \tfrac{1}{6} \, \mexp^0_{ij} \mathscr{R} \big) \, .\label{g2}
\end{align}
The right hand side is a matter modified form of the Schouten tensor. 
From this expression we immediately deduce the traces
\begin{align}
	t^{(2)} \ =& \ - \tfrac{1}{6} \mathscr{R}~, \qquad \quad 
	t^{(2,2)} \ =  \ \tfrac{1}{4} \big( \mathscr{R}_{ij} \mathscr{R}^{ij} - \tfrac{2}{9} \mathscr{R}^2 \big) ~.
\end{align}
The $zz$ component of the Einstein equation in \eqref{geom}, together with \eqref{Rzz}, determines the traces of higher order components in the expansion of the bulk metric:
\begin{align}
	u^{(0)} \ =& \ - 2 \Xb^2~, \\
	u^{(1)} \ =& \ - 4 \Xb \Xs + \tfrac{1}{96} \left( \langle b^+, b^-_2 \rangle_{\mexp^0} + \langle b^-, b^+_2 \rangle_{\mexp^0} \right) ~,\\
	 4 t^{(4)}  \ =& \ \ t^{(2,2)} - u^{(0)} - 3 u^{(1)} - 3 \Xb^2 - 8 \Xs^2 - 12 \Xb \Xs + \tfrac{1}{12} \left( \langle \mf, \mf \rangle_{\mexp^0} + \tfrac{1}{2}\langle F^I, F^I \rangle_{\mexp^0} \right) \nn \\
	&\ - \tfrac{1}{6} \langle {b}_1^+ , {b}_1^- \rangle_{\mexp^0} - \tfrac{1}{12} \langle b^- , ( \mexp^2 \circ b^+ ) \rangle_{\mexp^0} + \tfrac{1}{24} \left( \langle b^+, b^-_3 \rangle_{\mexp^0} + \langle b^-, b^+_3 \rangle_{\mexp^0} \right) \, .
\end{align}
Returning to the $ij$ component we may determine the logarithmic terms in \eqref{metricexp}:
\begin{align}
	h^0_{ij} \ =& \ \tfrac{1}{4} \mexp^0_{ij} ( u^{(0)} + 2 u^{(1)} + 8 \Xb \Xs ) \nn \\
	& \ - \tfrac{1}{16} \Big[ (b^+)_{(i}{}^k ( b_2^- )_{j)k} + (b^-)_{(i}{}^k ( b_2^+ )_{j)k} - \tfrac{1}{6} \mexp^0_{ij} \big( \langle b^+ , b_2^- \rangle_{\mexp^0} + \langle b^- , b^+_2 \rangle_{\mexp^0} \big) \Big] ~, \label{h0eqn}\\[5pt]
	h^1_{ij} \ =& \ - \tfrac{1}{2} h^0_{ij} + \mexp^2_{ik} ( \mexp^0 )^{kl} \mexp^2_{lj} + \tfrac{1}{4} \mexp^0_{ij} \big( 4 t^{(4)} - 2 t^{(2,2)} + u^{(1)} + 8 \Xs^2 \big) \nn \\
	& \ + \tfrac{1}{4} \big( \nabla^k \nabla_i \mexp^2_{jk} + \nabla^k \nabla_j \mexp^2_{ik} - \nabla^2 \mexp^2_{ij} - \nabla_i \nabla_j t^{(2)} \big)- \tfrac{1}{8} \big( (b^+_1)_{(i} (b^-_1)_{j)} - \tfrac{1}{3} \mexp^0_{ij} \langle b^+_1 , b^-_1 \rangle_{\mexp^0} \big)  \nn \\
	&\ + \tfrac{1}{8} \big[ (b^-)_{(i|k|} ( \mexp^2 )^{kl} (b^+)_{j)l} - \tfrac{1}{3} \mexp^0_{ij} (b^-)_{k}{}^m ( \mexp^2 )^{kl} (b^+)_{lm} \big] \nn \\
	&\ - \tfrac{1}{8} \big[ (b^+)_{(i}{}^k ( b_3^- )_{j)k} + (b^-)_{(i}{}^k ( b_3^+ )_{j)k} - \tfrac{1}{6} \mexp^0_{ij} \big( \langle b^+ , b_3^- \rangle_{\mexp^0} + \langle b^- , b^+_3 \rangle_{\mexp^0} \big) \big] \nn \\
	&\ - \tfrac{1}{4} \big[\mf_{ik} \mf_{j}{}^k + \tfrac{1}{2}F^I_{ik} F^I_{j}{}^k - \tfrac{1}{6} \mexp^0_{ij} \big( \langle \mf, \mf \rangle_{\mexp^0} + \tfrac{1}{2}\langle F^I, F^I \rangle_{\mexp^0} \big) \big] \, .\label{h1eqn}
\end{align}
The structure of the $ij$ component of the Einstein equation in four dimensions is such that $\mexp^4$ always appears with zero coefficient, and so is left undetermined. In the original literature \cite{deHaro:2000vlm} the $iz$ component has been used to determine $\mexp^4$ up to an arbitrary symmetric divergence-free tensor. However, in the supergravity we are considering the presence of a $(\log z)^2$ contribution to the bulk scalar field expansion means that $\Xs$ appears without a derivative, which hence spoils this approach. In section \ref{SecSUSYexpand} we will see that by imposing supersymmetry we obtain further constraints on the fields, and in particular this leads to an expression for $\mexp^4$ in terms of other 
data.

\subsection{Holographic renormalization}\label{SecHoloRenormalization}

Having solved the bulk equations of motion to the relevant order, we are now in a position to holographically renormalize
the Euclidean Romans theory. The bulk action~(\ref{IEuclid}) is divergent for 
an asymptotically locally hyperbolic solution, but can be rendered finite by the addition of appropriate local 
counterterms. The corresponding computations in Lorentzian signature have been carried out in \cite{Ohl:2010au}. 

We begin by taking the trace of the Einstein equation \eqref{geom}. Substituting the result together with \eqref{Beom}  into the Euclidean action \eqref{IEuclid}, we  arrive at the bulk on-shell action
\begin{align}
	I_{\text{on-shell}} \ = \ \frac{1}{2\kappa_5^2} \int_{\B_5} \ & \Big[\tfrac{8}{3} ( X^2 + 2 X^{-1} ) \, {*1} +\tfrac{1}{3} X^4 \cF \wedge * \cF + \tfrac{1}{6} X^{-2} \cF^I \wedge * \cF^I \nn \\
	& -\tfrac{1}{12} X^{-2} {\mathcal{B}^-} \wedge * \mathcal{B}^+ +\tfrac{\ii}{4} \cF^I \wedge \cF^I \wedge \cA \Big]\, . \label{IEuclidOnShell}
\end{align}
Here $\B_5$ is the bulk five-manifold, with boundary $\partial \B_5=M_4$. In order to obtain the equations of motion  (\ref{Xeom})--(\ref{geom})
from the original bulk action (\ref{IEuclid}) on a manifold with boundary, one has to add the  Gibbons-Hawking term
\begin{align}
	I_{\text{GH}} \ =& \ - \frac{1}{\kappa_5^2} \int_{\partial \B_5} \dd^4 x \, \sqrt{\det h} \, K \ = \ \frac{1}{\kappa_5^2} \int_{\partial \B_5} \dd^4 x \, {z} \partial_z \sqrt{\det h} \, .
\end{align}
Here, more precisely, one cuts $\B_5$ off at some finite radial distance, or equivalently non-zero $z>0$, 
and $(M_4,h)$ is the resulting four-manifold boundary, with trace of the second fundamental form being $K$. 
Recall from (\ref{FGmetric}) that $h_{ij} = \frac{1}{z^2} \mexp_{ij}$.

The combined action $I_{\text{on-shell}}+I_{\text{GH}}$ suffers from divergences as the conformal boundary is approached. To remove these divergences we use the standard method of holographic  renormalization \cite{Emparan:1999pm,Taylor:2000xw,deHaro:2000vlm}. Namely, we introduce a small cut-off $z=\cutoff>0$, and expand all fields via the Fefferman-Graham expansion of section \ref{SecFG} to identify the divergences. 
These may be cancelled by adding local boundary counterterms. We find
\begin{align}
	I_{\text{counterterm}} \ = \ 
	\frac{1}{\kappa_5^2} \int_{\partial \B_5} &\dd^4 x \, \sqrt{\det h}\, \Big\{  {3} +\tfrac{1}{4} R ( h ) +{3} ( X - 1 )^2 - \tfrac{1}{32} \langle {\mathcal{B}^-} , \mathcal{B}^+ \rangle_h \nn \\
	&+\log \cutoff \, \Big[- \tfrac{1}{8} \Big( \mathscr{R}_{ij} ( h ) \mathscr{R}^{ij} ( h ) - \tfrac{1}{3} \mathscr{R}(h)^2 \Big)  +\tfrac{3}{2} ( \log \cutoff )^{-2} ( X - 1 )^2\nn \\
	&\qquad \qquad +\tfrac{1}{48} \langle {H}^- , H^+ \rangle_h +\tfrac{1}{8} \langle \mathcal{F} , \mathcal{F} \rangle_h + \tfrac{1}{16} \langle \cF^I , \cF^I \rangle_h \Big]\Big\} \, .\label{ctaction}
\end{align}
Notice the somewhat unusual form of the logarithmic term for the scalar field $X$, but {\it c.f.} the 
expansion (\ref{Xexp}). 
As is  standard, we have written the counterterm action (\ref{ctaction}) covariantly in terms of the induced metric $h_{ij}$ on $M_4=\partial \B_5$.
The total renormalized action is then
\beq
S \ = \ \lim_{\cutoff\rightarrow 0} \, \left(I_{\text{on-shell}} + I_{\text{GH}}+I_{\text{counterterm}}\right)~,\label{Actionfinal}
\eeq
which by construction is finite.

The choice of local counterterms (\ref{ctaction}) defines a particular renormalization scheme, that is in some sense a ``minimal scheme'' 
in the case at hand. However, we are free to consider a non-minimal scheme where we add local counterterms to the action which remain finite as $\cutoff \rightarrow 0$. For the supergravity theory we are considering, the following are an independent set of finite counterterms that 
are both diffeomorphism and gauge invariant:\footnote{We may also add finite local counterterms constructed from the $B$-field. 
For example, terms proportional to $\int_{\partial\B_5}\diff^4 x \sqrt{\det h}\, \langle H^-, H^+\rangle_h$, or $\int_{\partial\B_5}\diff^4 x \sqrt{\det h}\, R(h)\langle \mathcal{B}^-, \mathcal{B}^+\rangle_h$. However, for the topological twist we will later set the $B$-field to zero, and these 
terms will not be relevant to our discussion.}
\begin{align}
	I_{\text{ct,\, finite}} \, = \, - \frac{1}{\kappa_5^2} \int_{\partial \B_5} \dd^4 x \, \sqrt{\det h} \, \Big[ &\zeta_1 R^2 + \zeta_2 C_{ijkl} C^{ijkl} + \zeta_3 \mathcal{F}_{ij} \mathcal{F}^{ij} + \zeta_4 \cF^I_{ij} \cF^{I ij}\nn \\
 	&+\zeta_5 \mathcal{E} + \zeta_6 \mathcal{P}  + \zeta_7 \epsilon^{ijkl}  \mathcal{F}_{ij} \mathcal{F}_{kl} + \zeta_8 \epsilon^{ijkl} \cF^I_{ij} \cF^I_{kl}\Big] \, .\label{Ifinite}
\end{align}
Here $\zeta_1, \ldots , \zeta_8$ are arbitrary constant coefficients, $C_{ijkl}$ denotes the Weyl tensor of the metric $h_{ij}$, 
while the Euler scalar $\mathcal{E}$ and Pontryagin scalar $\mathcal{P}$ are respectively
\begin{align}
	%
	\mathcal{E} \ = & \ R_{ijkl} R^{ijkl} - 4 R_{ij} R^{ij} + R^2 \, , \qquad
	\mathcal{P} \ = \ \tfrac{1}{2} \epsilon^{ijkl} R_{ijmn} R_{kl}{}^{mn} \, .\label{EandP}
\end{align}
In particular, notice that for compact $M_4=\partial \B_5$ without boundary, the second line of (\ref{Ifinite}) are all topological invariants: they are  proportional to the Euler number $\chi(M_4)$, 
the signature $\sigma(M_4)$, and the Chern numbers $\int_{M_4}c_1(\mathcal{L})^2$, $\int_{M_4}c_2(\mathcal{V})$ respectively, where 
$\mathcal{L}$ and $\mathcal{V}$ denote the rank 1 and rank 2 complex vector bundles associated to the $U(1)_R$ and 
$SU(2)_R$ gauge bundles, respectively. In the real Euclidean theory in which we are working, recall that $\mathcal{F}=\diff \mathcal{A}$ is globally exact 
(and purely imaginary), and in any case for the topological twist studied later in the paper we will have $\mathcal{A}\mid_{M_4}=0$. 
Being topological invariants, the variation of the action we shall compute in section \ref{SecVary} will be insensitive to 
the choice of constants $\zeta_5,\ldots,\zeta_8$. 

As emphasized in \cite{Genolini:2016ecx}, in order 
to make quantitative comparisons in AdS/CFT it is important to match choices of renormalization schemes 
on the two sides. In particular, localization calculations in QFT make a (somewhat implicit) choice of scheme. In the case at hand, we note 
that in \cite{Crossley:2014oea} a supersymmetric R\'enyi entropy, computed in field theory using localization, was 
successfully matched to a gravity calculation involving a supersymmetric black hole in the $\mathcal{N}=4^+$ Romans theory. 
Here the supergravity action was computed using the minimal scheme. Our computation in section \ref{SecVary} 
will imply that this minimal scheme is indeed the correct one to compare to the topological twist of \cite{Witten:1988ze}. 
We shall make further comments on this, and the relation to recent papers \cite{Genolini:2016sxe, Genolini:2016ecx, Papadimitriou:2017kzw, An:2017ihs}, in section \ref{SecProof}.

Given the renormalized action we may compute the following VEVs:
\begin{align}
	\langle T_{ij} \rangle \ =& \ \frac{2}{\sqrt{g}} \frac{\delta S}{ \delta g^{ij} } \, , \, \, \qquad \langle \Xi \rangle \ = \ \frac{1}{\sqrt{g}} \frac{\delta S}{ \delta \Xb } \, , \nn\\  \langle \Jcur_I^{i} \rangle \ = & \ \frac{1}{\sqrt{g}} \frac{\delta S}{ \delta A^I_i }~, 
	\qquad \langle \, \JRcur^i \rangle \ = \ \frac{1}{\sqrt{g}} \frac{\delta S}{ \delta \ma_i }~.
	\label{VEVs}
\end{align}
Here, as usual in AdS/CFT, the boundary fields $\mexp^0_{ij}=g_{ij}$, $\Xb$, $A^I_i$ and $\ma_i$ act as sources for operators, and 
the expressions in (\ref{VEVs}) compute the vacuum expectation values of these operators. Similar 
expressions may also be written for the boundary fields  $b^\pm$ for $\mathcal{B}^\pm$, but these will be zero for the topological twist of interest 
and play no role in the present paper.  Using 
the above holographic renormalization we may write (\ref{VEVs}) as the following limits:
\begin{align}
	\langle T_{ij} \rangle \ = \ \frac{1}{\kappa_5^2} &\lim_{\cutoff \to 0} \, \frac{1}{\cutoff^2} \bigg[ - K_{ij} + K h_{ij} - \big( 3 + 3(X-1)^2 \big) h_{ij} + \tfrac{1}{2} \left( \mathscr{R}_{ij} (h) - \tfrac{1}{2} \mathscr{R}(h)\, h_{ij} \right) \nn \\
	&\qquad \quad \ \ - \tfrac{1}{32} \langle \mathcal{B}^- , \mathcal{B}^+ \rangle_h \, h_{ij} \nn \\
	&+ \log \cutoff \bigg( \tfrac{1}{4} \mathscr{B}_{ij}(h) + \tfrac{1}{2} \mathcal{F}_{ik}\mathcal{F}_j{}^k - \tfrac{1}{8} h_{ij} \langle \mathcal{F} , \mathcal{F} \rangle_h + \tfrac{1}{4} \cF^I_{ik} \cF^I_j{}^k - \tfrac{1}{16} h_{ij} \langle \cF^I , \cF^I \rangle_h \nn \\
	&+ \tfrac{1}{8} {H}^-_{ikl} {H^+}_j{}^{ kl} - \tfrac{1}{48} h_{ij} \langle {H}^- , H^+ \rangle_h - \tfrac{3}{2} ( \log \cutoff )^{-2} ( X - 1 )^2 h_{ij} \bigg) \bigg]~, 
\end{align}
where $K_{ij}$ is the second fundamental form of the cut-off hypersurface $(M_4,h_{ij}$) and
the $B$-field modified Bach tensor is ({\it c.f}.\ (\ref{BRicci}))
\begin{align}
	\mathscr{B}_{ij} \ =& \ - \tfrac{2}{3} \nabla_i \nabla_j \mathscr{R} - \nabla^2 \Big( \mathscr{R}_{ij} - \tfrac{1}{6} h_{ij} \mathscr{R} \Big) + 2 \nabla_k \nabla_{(i} \mathscr{R}^k{}_{j)} - 2 \mathscr{R}_{ik} \mathscr{R}^k{}_j + \tfrac{2}{3} \mathscr{R} \mathscr{R}_{ij} \nn \\
	& \ + \tfrac{1}{2} h_{ij} \Big( \mathscr{R}_{kl} \mathscr{R}^{kl} - \tfrac{1}{3} \mathscr{R}^2 \Big) \, ,
\end{align}
together with
\begin{align}
	\langle \Xi \rangle \ =& \ \frac{1}{\kappa_5^2} \lim_{\cutoff \to 0} \, \frac{\log \cutoff}{\cutoff^2} \bigg[ - 3 X^{-2} \cutoff \partial_\cutoff X + 6 ( X - 1 ) + 3 ( \log \cutoff )^{-1} ( X - 1 ) \bigg] ~,\nn \\
	\langle \Jcur^{Ii} \rangle \ =& \ \frac{1}{4\kappa_5^2} \lim_{\cutoff \to 0} \, \frac{1}{\cutoff^4} \bigg\{- *_h \Big[ \dd x^i \wedge ( X^{-2} *_5 \cF^I +  \ii \cF^I \wedge \mathcal{A} ) \Big] + \log \cutoff \, \mathcal{D}_j \cF^{Iij} \bigg\}~,\nn\\
	\langle\,  \JRcur^{i} \rangle \ =& \ \frac{1}{2\kappa_5^2} \lim_{\cutoff \to 0} \, \frac{1}{\cutoff^4} \bigg[- *_h \Big( \dd x^i \wedge  X^{4} *_5 \cF  \Big) + \log \cutoff \, \nabla_j \cF^{ij} \bigg]~. 
\end{align}
Here $*_h$ denotes the Hodge duality operator for the  metric $h_{ij}$. A computation then gives the finite expressions
\begin{align}
	\langle T_{ij} \rangle \ =& \   \frac{1}{\kappa_5^2} \bigg[ 2 \mexp^4_{ij} + \tfrac{1}{2} h^1_{ij} - \tfrac{1}{2} ( 4 t^{(4)} - 2 t^{(2,2)} - \tfrac{1}{2} u^{(1)}  ) \mexp_{ij}^{0}  - 3 \mexp_{ij}^{0} \Xs^2 - \mexp_{ij}^{2} t^{(2)}\nn \\
	&\qquad + \tfrac{1}{4} \Big( \nabla^k \nabla_i \mexp^2_{jk} + \nabla^k \nabla_j \mexp^2_{ik} - \nabla^2 \mexp^2_{ij} - \nabla_i \nabla_j t^{(2)} \Big) + \tfrac{1}{4} \mexp_{ij}^{0} \big( \mexp^2_{kl} \mathscr{R}^{kl} \big) - \tfrac{1}{4} \mexp_{ij}^{2} R
	 \nn \\
	&\qquad - \tfrac{1}{8} \big[ (b^+)_{(i}{}^k (b^-_3)_{j)k} + (b^-)_{(i}{}^k (b^+_3)_{j)k} - \tfrac{1}{2} \mexp^0_{ij} \big(  \langle {b}^+ , {b}_3^- \rangle_{\mexp^0} + \langle {b}^- , {b}_3^+ \rangle_{\mexp^0} \big) \big] \nn \\
	&\qquad + \tfrac{1}{8} \big[ (b^+)_{(i|k|} (\mexp^2)^{kl} ({b}^-)_{j)l} - \tfrac{1}{2} \mexp^0_{ij} \langle b^- , ( \mexp^2 \circ b^+ ) \rangle_{\mexp^0} \big] \nn \\
	&\qquad - \tfrac{1}{32} \mexp_{ij}^{0}  \big[  \langle {b}^+ , {b}_3^- \rangle_{\mexp^0} + \langle {b}^- , {b}_3^+ \rangle_{\mexp^0} - 2 \langle b^- , ( \mexp^2 \circ b^+ ) \rangle_{\mexp^0} \big] \nn \\
	&\qquad - \tfrac{1}{4} \mexp_{ij}^{0} \big( \nabla^k \nabla^l \mexp^2_{kl} - \nabla^2 t^{(2)} \big) - \tfrac{1}{128} \mexp_{ij}^{0} \big(  \langle {b}^+ , {b}_2^- \rangle_{\mexp^0} + \langle {b}^- , {b}_2^+ \rangle_{\mexp^0} \big) \bigg]  \, ,\label{Tij} \\
	\langle \Xi \rangle \ =& \  \ \frac{3}{\kappa_5^2} \Xs \, , \label{Xi}\\
	\langle \Jcur^{I}_i \rangle \ =&  \ -\frac{1}{4\kappa_5^2}\left[(\aIfour)_i+2(\aIfive)_i - \ii \big( *_4(\ma\wedge F^I) \big)_i \right]\label{calJi}~,\\
	\langle \, \JRcur_i \rangle \ =&  \ -\frac{1}{2\kappa_5^2}\left[(\mafour)_i+2(\mafive)_i \right]\label{bbJ}~.
\end{align}
Notice that these expressions contain a number of terms that are not determined, in terms of boundary data, by the Fefferman-Graham expansion of the bosonic equations of motion. In particular the $\mexp^4_{ij}$ term in the stress-energy tensor $T_{ij}$, the scalar $\Xs$ that determines $\Xi $, and $\aIfive$, $\mafive$ 
 appearing in the $SU(2)_R$ and $U(1)_R$ current, respectively. The general holographic Ward identity corresponding to the first three variations of the 
 action is given by equation~(\ref{holWard}). We will need the expressions (\ref{Tij})--(\ref{calJi}) in section \ref{SecVary}.


\section{Supersymmetric solutions}\label{SecSUSY}

In this section we study supersymmetric solutions to the Euclidean $\mathcal{N}=4^+$ theory. We begin in section \ref{SecBoundaryKSE} 
by deriving the Killing spinor equations on the conformal boundary, starting from the bulk equations (\ref{BulkGravitino}), 
(\ref{BulkDilatino}). We precisely recover the Euclidean $\mathcal{N}=2$ conformal supergravity equations of \cite{Klare:2013dka}. 
 In section \ref{SecTT} we then recall from \cite{Karlhede:1988ax} how the topological twist arises as a special 
 solution to these Killing spinor equations, that exists on any Riemannian four-manifold $(M_4,g)$. We rephrase this in terms 
 of the quaternionic K\"ahler structure that exists on any such manifold, involving (locally) a triplet of self-dual two-forms $\Jb^I$.  
Finally, in section \ref{SecSUSYexpand} we expand solutions to the bulk spinor equations in a Fefferman-Graham-like 
expansion.

\subsection{Boundary spinor equations}\label{SecBoundaryKSE}

We begin by expanding the bulk Killing spinor equations  (\ref{BulkGravitino}), 
(\ref{BulkDilatino}) to leading order near the conformal boundary at $z=0$. We will consequently need the 
Fefferman-Graham expansion of an orthonormal frame for the metric (\ref{FGmetric}), (\ref{metricexp}), together with the associated spin connection. The following is a choice of frame ${\rm E}_\mu^{\overline{\mu}}$ for the metric (\ref{FGmetric}):
\begin{equation}
{\rm E}^{\overline{z}}_z \ = \ \frac{1}{z}, \qquad {\rm E}^{\overline{z}}_i \ = \ {\rm E}^{\overline{i}}_z \ =  \ 0, \qquad  {\rm E}_i^{\overline{i}} \ = \ \frac{1}{z}\te^{\overline{i}}_i~,\label{FGframe}
\end{equation}
where $\te^{\overline{i}}_i$ is a frame for the $z$-dependent metric $\mexp$. The latter then has the expansion (\ref{metricexp}), but for the 
present subsection we shall only need that
\begin{equation}
\te^{\overline{i}}_i \ = \ \ex^{\overline{i}}_i + O(z^2)~,
\end{equation}
where $ \ex^{\overline{i}}_i$ is a frame for the boundary metric $\mexp^0=g$.
The non-zero components of the  spin connection $\Omega_\mu^{\ \overline{\nu}\overline{\rho}}$ at this order 
are correspondingly
\begin{align}
 \Omega_{{i}}^{\ \overline{zj}} \ = \ \frac{1}{z}\ex_{i}^{\, \overline{j}}+O(z)~, \qquad \Omega_{{i}}^{\ \overline{j k}} \ =\  (\omega^{(0)})_{{i}}^{\ \overline{jk}} + O(z^2)~,
\end{align}
where $(\omega^{(0)})_{{i}}^{\ \overline{j k}}$ denotes the boundary spin connection. 

The generators $\gamma_{\bar{\mu}}$ of the Clifford algebra $\mathrm{Cliff}(5,0)$ in this frame are chosen to obey 
\begin{equation}
\gamma_{\bar{z}} \ = \  \gamma_{\bar{1}\bar{2}\bar{3}\bar{4}}~.
\end{equation}
It follows that $\gamma_{\bar{z}}^2=1$, 
and we may identify $-\gamma_{\bar{z}}$ with the boundary chirality operator. The bulk Killing spinor is then expanded as
\begin{equation}
	\epsilon \ = \ z^{-1/2}\los + z^{1/2}\eta + o(z^{1/2})~.
\end{equation}
As in (\ref{pmeigen}), we may further decompose the spinors $\los$, $\eta$ into their projections 
$\los^\pm$, $\eta^\pm$ onto the $\pm \ii$ eigenspaces 
of $\Gamma_{45}$. At leading order in the $z$-component of the gravitino equation (\ref{BulkGravitino}) one then finds
\begin{equation}
	- \gamma_{\bar{z}} \los^\pm \ = \ \pm \los^\pm \, ,\label{loschiral}
\end{equation}
so that the $\Gamma_{45}$ eigenvalue of the leading order spinor $\los$ is correlated with its boundary chirality. 
Similarly, at the next order in the gravitino equation one finds the opposite correlation for the spinor $\eta$:
\begin{equation}
	- \gamma_{\bar{z}} \eta^\pm \ = \ \mp \eta^\pm \, .\label{etachiral}
\end{equation}

Recall that the boundary $B$-fields satisfy $*_4 b^\pm = \mp b^\pm$ (see (\ref{b0Duality})). This together with the chirality 
conditions (\ref{loschiral}) implies that
\begin{equation}
	b^\pm \cdot \los^\pm \ = \ 0 \, ,
\end{equation}
where $\cdot$ denotes the Clifford product (using the boundary frame). Using this, the leading order term in the $i$-component of the gravitino equation is then seen to be identically satisfied. The next order gives the pair of boundary Killing spinor equations:
\begin{equation}
	 \mathcal{D}^{(0)}_i\los^\pm  - 
	\tfrac{\ii}{4}b^\mp_{{i}{j}}\gamma^{j}\los^\mp \mp \gamma_{i}\eta^\pm \ = \ 0~, \label{BdyKSEs}
\end{equation}
where we have defined the 
covariant derivative
\begin{equation}
\mathcal{D}^{(0)}_i \ \equiv \  \nabla^{(0)}_{{i}} \pm \tfrac{\ii}{2}\ma_{{i}} + \tfrac{\ii}{{2}} A_{{i}}^I\sigma_I~.
\end{equation}
Here $\nabla^{(0)}_{i}$ denotes the Levi-Civita spin connection of the boundary metric $\mexp^0_{ij}=g_{ij}$, and $\gamma_i = \gamma_{\bar{i}}\, \ex^{\bar{i}}_i$, so that $\{\gamma_i,\gamma_j\}=2g_{ij}$.

Turning to the bulk dilatino equation (\ref{BulkDilatino}), the leading order term is in fact equivalent to the duality properties of $b^\pm$, given the chiralities of $\los^\pm$. At the next order we obtain the boundary dilatino equation
\begin{equation}
	-\mathrm{f}\cdot \los^\pm \pm \tfrac{1}{{2}}F^I\sigma_I\cdot \los^\pm \mp 3\ii \Xb\, \los^\pm + \tfrac{1}{2} b^\mp \cdot\eta^\mp \mp \tfrac{1}{2}b_1^\mp \cdot \los^\mp \ = \ 0~.\label{BdyDils}
\end{equation}

The supersymmetry equations for four-dimensional Euclidean off-shell $\mathcal{N}=2$ conformal supergravity have been studied\footnote{See \cite{Gupta:2012cy} for related earlier work and \cite{deWit:2017cle} for a recent construction of Euclidean $\mathcal{N}=2$ conformal supergravity from a timelike reduction of a five-dimensional theory.} in \cite{Klare:2013dka}, 
and our equations (\ref{BdyKSEs}), (\ref{BdyDils}) precisely reproduce the equations in this reference.\footnote{The explicit 
notation change is  $A_4^{\mathrm{KZ}}=-\ii\ma  $, $A^I_{\mathrm{KZ}}=A^I$, $ T^\pm_{\mathrm{KZ}}=-b^\pm $, $\epsilon_\pm^{\mathrm{KZ}}=\los^\mp$, $\tilde{d}_{\mathrm{KZ}}=2\Xb$.} Notice in particular 
that one can solve for the (conformal) spinor $\eta$ by taking the trace of (\ref{BdyKSEs}) with $\gamma^i$, to obtain
\begin{equation}
\eta^\pm \ = \ \pm \tfrac{1}{4}\sla{\mathcal{D}^{(0)}}\los^\pm~,
\end{equation}
where $\sla{\mathcal{D}^{(0)}} \equiv \gamma^{{i}}\mathcal{D}^{(0)}_{{i}}$ is the Dirac operator. 
Taking the covariant derivative of (\ref{BdyKSEs}) and using the integrability condition for $[\mathcal{D}^{(0)}_i,\mathcal{D}^{(0)}_j]$ 
then leads to the following form of the dilatino equation
\begin{equation}
\sla{\mathcal{D}^{(0)}}\sla{\mathcal{D}^{(0)}}\los^\pm - \ii \mathcal{D}_{{i}}(b^\mp)^{{i}}_{\ {j}}\gamma^{{j}}\los^\mp 
+ \left(4\Xb + \tfrac{1}{3}R\right)\los^\pm \mp 2\ii\,  \mf \cdot \los^\pm \ =\  0~,\label{BdyDilagain}
\end{equation}
where $R=R(g)$ is the Ricci scalar of the boundary metric.
Requiring the boundary 
fields $g_{ij}, \Xb, \ma, A^I,b^\pm$ to solve the spinor equations (\ref{BdyKSEs}), (\ref{BdyDils}) for 
$\los^\pm$ in general imposes geometric constraints. Remarkably, in \cite{Klare:2013dka} 
it is shown that generically these conditions are equivalent to the boundary manifold 
$(M_4,g)$ admitting a conformal Killing vector. However, the topological twist 
background of \cite{Karlhede:1988ax} arises as a very degenerate case, where in fact
 $(M_4,g)$ may be an arbitrary Riemannian four-manifold. We turn to this case in the next subsection.

\subsection{Topological twist}\label{SecTT}

The topological twist background of \cite{Karlhede:1988ax} is obtained by setting 
\begin{equation}
\los^- \ = \ 0~, \qquad \ma \ = \ 0~, \qquad b^\pm \ = \ 0~, \qquad \eta^\pm \ = \ 0~.
\end{equation}
The boundary Killing spinor equation (\ref{BdyKSEs}) immediately implies that 
 $\los^+$ is covariantly constant
 \begin{equation}
 \mathcal{D}^{(0)}_i\los^+ \ =\ 0~.\label{Dlos}
 \end{equation}
The dilatino equation, in the form (\ref{BdyDilagain}), then fixes
\begin{equation}
\Xb \ = \  -\tfrac{1}{12}R~.\label{X2twist}
\end{equation}

Recall  that $\los^+$ is a doublet of positive chirality spinors: the Pauli matrices $\sigma_I$ act on these doublet indices, while the Clifford matrices $\gamma_{\bar{i}}$ act on the spinor indices. We may write out the covariant derivative in  (\ref{Dlos})  more explicitly by first 
 introducing the following explicit Hermitian representation
\begin{align}
\gamma_{\bar{a}} \ = \  \left(\begin{matrix} 0 & \ii \sigma_{\bar{a}} \\ -\ii \sigma_{\bar{a}} & 0\end{matrix}\right)~, \quad 
\gamma_{\bar{4}} \ = \ \left(\begin{matrix} 0 & -1_2 \\ -1_2 & 0\end{matrix}\right)~, \quad 
\gamma_{\bar{z}} \ =\   \left(\begin{matrix} 1_2 & 0 \\ 0 & -1_2\end{matrix}\right)~.\label{gammamatrices}
\end{align}
Here $\bar{a}=1,2,3$. Since $\gamma_{\bar{z}}\los^+=-\los^+$, we may identify each of the two spinors in the doublet $\los^+$ with a two-component spinor, acted 
on by the second $2\times 2$ block. With these choices (\ref{Dlos})  reads
 \begin{equation}
\mathcal{D}^{(0)}_{i}\los^+ \ =\  \partial_{{i}}\los^+ +\tfrac{\ii}{4}\eta^{\bar{a}}_{\overline{jk}}\, (\omega^{(0)})_{{i}}^{\ \overline{jk}} \sigma_{\bar{a}}\los^+ + \tfrac{\ii}{2}A_{{i}}^I\sigma_I\los^+ \ = \ 0~,\label{Dlosexplicit}
\end{equation}
where $\eta^{\bar{a}}_{\overline{ij}}$ are the self-dual 't Hooft symbols, and recall that 
$(\omega^{(0)})_{{i}}^{\ \overline{jk}}$ is the spin connection for the boundary metric $g_{ij}$. One may then solve (\ref{Dlosexplicit}) by taking
\beq
A^I_i \ =\  \tfrac{1}{2}\eta^{I}_{\overline{jk}}\, (\omega^{(0)})_{{i}}^{\ \overline{jk}}~, \qquad (\los^+)^i_{\ \alpha} \ = \ (\ii\sigma_2)^i_{\ \alpha}  \, c~.\label{TTsoln}
\eeq
Here $i=1,2$ labels the doublet indices, while $\alpha=1,2$ labels the positive chirality spinor indices, and notice that the 
frame index $\bar{a}=1,2,3$ is identified with the gauge indices $I=1,2,3$. It is straightforward to check that 
(\ref{TTsoln}) solves (\ref{Dlosexplicit}), for any constant $c$. The $SU(2)_R$ gauge field $A^I$ given by 
(\ref{TTsoln}) is precisely the right-handed part of the spin connection, where recall that 
$Spin(4)=SU(2)_-\times SU(2)_+$. Thus the $SU(2)_R$ gauge bundle is identified with $SU(2)_+$.

More invariantly, $\los^+$ is a section of $\mathcal{S}^+\otimes \mathcal{V}$, where 
$\mathcal{S}^+$ denotes the positive chirality spinor bundle over $M_4$, while $\mathcal{V}$ is the rank 2 complex vector 
bundle for which $A^I$ is an associated $SU(2)$ connection. {\it A priori} this makes sense globally only when $M_4$ is a spin manifold, when
$\mathcal{S}^+$ and $\mathcal{V}$ both exist as genuine vector bundles. However, the topological twist 
(\ref{TTsoln}) identifies $\mathcal{V}$ with $\mathcal{S}^+$, and their tensor product then always exists globally, 
even when $M_4$ is not spin.\footnote{There are various ways to see this. For example, the lack of a spin structure on $M_4$ 
is detected by a non-zero second Stiefel-Whitney class $w_2(M_4)\in H^2(M_4,\Z_2)$. Concretely this 
means the cocycle condition for the spin lift of the frame bundle fails up to some minus signs. However, 
if two copies are tensored together all such signs square to $+1$, and the tensor product is a well-defined bundle.}
This topological construction of a spin-type bundle on a manifold which is not necessarily spin
was first suggested in \cite{Back:1978zf}, and  
 is sometimes referred to as a $Spin_{\mathscr{G}}$ structure, where here the group $\mathscr{G}=SU(2)$. 
  Perhaps more familiar are 
$Spin^c$ structures, where instead $\mathscr{G}=U(1)$. (For example, this arises in Seiberg-Witten theory.)

It will be convenient later to introduce the triplet of self-dual two-forms
\begin{equation}
\Jb^I_{ij} \ \equiv \ \eta^{I}_{\overline{ij}}\,  \ex^{\overline{i}}_i \, \ex^{\overline{j}}_j~,\label{JItHooft}
\end{equation}
where recall that $\ex^{\overline{i}}_i$ is the boundary frame for $g_{ij}$. More explicitly, these read
\begin{align}
\Jb^1 \ =  \ \ex^2\wedge \ex^3+\ex^1\wedge \ex^4\, ,\quad
\Jb^2 \ =  \   \ex^3\wedge  \ex^1+\ex^2\wedge  \ex^4\, ,\quad
\Jb^3 \ = \   \ex^1\wedge  \ex^2+\ex^3\wedge\ex^4~.\label{JIframe}
\end{align}
Of course, in general a frame $\ex^{\overline{i}}_i$ is only defined locally on $M_4$, in an appropriate open set, and likewise the $\Jb^I$ in (\ref{JIframe}) 
are then well-defined forms only locally. More globally, local frames are patched together with $SO(4)$. The spin cover is $Spin(4)\cong 
SU(2)_-\times SU(2)_+$, and the self-dual/anti-self-dual two-forms are precisely the representations associated to $SO(3)_\pm = SU(2)_\pm/\Z_2$. 
In particular, the $\{\Jb^I\}$ rotate as a 3-vector under $SO(3)_+\subset SO(4)$. In this sense the $\Jb^I$ in general don't exist 
individually as global two-forms on $M_4$, but instead as a triplet of forms that rotate appropriately. We comment further on this below.

One can also write the $\Jb^I$ in terms of spinor bilinears. 
Recall from the end of section~\ref{SecRomans} that the bulk spinors satisfy a symplectic Majorana reality condition. 
In particular the boundary spinor $\los^+$ satisfies 
\beq\label{losreal}
(\los^+)^c \ \equiv \ \ii \sigma_2  \mathscr{C}(\los^+)^* \ = \ \los^+~,
\eeq
where recall that $\mathscr{C}$ is the charge conjugation matrix for the spacetime Clifford algebra. In the explicit basis (\ref{gammamatrices}) we may take
\beq
\mathscr{C} \ =\ \begin{pmatrix} \, \ii\sigma_2 & 0 \\ 0 & \ii\sigma_2\end{pmatrix}~.
\eeq
Given the solution (\ref{TTsoln}) one finds that the reality condition (\ref{losreal}) is satisfied provided 
the constant $c\in \R$. Explicitly, the components of the doublet $\los^+$ are
\beq
(\los^+)^1 \ = \ (0,0,0,c)^{\mathrm{T}}~, \qquad (\los^+)^2 \ = \ (0,0,-c,0)^{\mathrm{T}}~.\label{spinorexpl}
\eeq
We then define the boundary spinor
\beq
\chi \ \equiv \ (\los^+)^1~.\label{definechi}
\eeq
This has square norm $\bar\chi\chi=c^2$, where the bar denotes Hermitian conjugate, and $\chi$ of course has positive chirality, $-\gamma_{\bar{z}}\chi = \chi$.  One easily checks that
\begin{align}
\Jb^2 + \ii \Jb^1 \ = \ \frac{1}{\bar\chi\chi}\, \bar{\chi}^c\gamma_{(2)}\chi~, \qquad \Jb^3 \ = \ \frac{\ii}{\bar\chi\chi} \, 
\bar\chi \gamma_{(2)}\chi~,\label{JIbilinears}
\end{align}
where $\chi^c\equiv \mathscr{C}\chi^*$. 

From the original definition (\ref{JItHooft}), the $\Jb^I$ inherit a number of algebraic identities from those for the 't Hooft symbols.
For example,
\beq
\Jb^I_{ij}\Jb^I_{kl} \ = \ g_{ik}g_{jl} - g_{il}g_{jk} + \epsilon_{ijkl}~.\label{outerJI}
\eeq
Using the metric to raise an index, one obtains a triplet $(\Ib^I)^i_{\ j}\equiv g^{ik}(\Jb^I)_{kj}$ of endomorphisms of the tangent bundle of $M_4$. 
These
satisfy the quaternionic algebra
\beq\label{Qua}
\Ib^I\circ \Ib^J \ =\ -\delta ^{IJ} -\epsilon^{IJ}_{\ \ K} \Ib^K~.
\eeq
One also finds that
\beq\label{QKeqn}
\nabla_i \Jb^I_{jk} \ = \  \epsilon^I_{\ JK} A^J_i \Jb^K_{jk}~,
\eeq
where the R-symmetry gauge field $A^I$ here is precisely the right-handed spin connection given by the topological twist 
(\ref{TTsoln}). Notice that we may correspondingly write the curvature as
\beq
F^I_{ij} \ = \ \tfrac{1}{2}\Jb^I_{kl} R_{ij}^{\ \ kl}~,\label{Ftwist}
\eeq
where $R_{ijkl}$ is the boundary Riemann tensor.

In general a \emph{quaternionic K\"ahler manifold} is a Riemannian manifold of dimension $4n$ with holonomy $Sp(n)\cdot Sp(1)\subset SO(4n)$.\footnote{See, for example, \cite{salamon}.} 
Such manifolds admit, locally, a triplet of skew endomorphisms $\Ib^I$ of the tangent bundle satisfying (\ref{Qua}), 
for which the corresponding triplet of two-forms $\Jb^I$ satisfy (\ref{QKeqn}). 
Here $A^I$ is the Riemannian connection corresponding  to the $Sp(1)$ part of this holonomy group. 
For $n=1$ notice that $Sp(1)\cdot Sp(1)=SO(4)$, and such a structure exists on any Riemannian 
four-manifold $(M_4,g)$ (as we have just seen). Crucially, the two-forms (\ref{JIframe}) are not in general defined globally, 
but  are (in our language)  twisted by the R-symmetry gauge field, transforming as a vector under $SO(3)_R=SU(2)_R/\Z_2$. 
As such, they don't define a reduction of the structure group to $SU(2)_-$, as a global set of such forms would do. 
Indeed, the globally defined tensor on a quaternionic K\"ahler manifold is the four-form 
$\Psi\equiv \Jb^I\wedge\Jb^I$ (summed over $I$), and in four dimensions ($n=1$) this is proportional to the volume form. 
The stabiliser of $\Psi$ is $Sp(n)\cdot Sp(1)$, which is $SO(4)$ when $n=1$.

In dimensions $n\geq 2$ irreducible quaternionic K\"ahler 
manifolds are automatically Einstein. Some authors choose to \emph{define} a quaternionic K\"ahler 
four-manifold to be an Einstein manifold with self-dual Weyl tensor, but we shall not use this terminology.

\subsection{\texorpdfstring{$U(1)_R$}{U(1)R} current}\label{SecU1current}

Before continuing to expand the spinor equations into the bulk, in this subsection we pause briefly 
to consider the VEV of the $U(1)_R$ current given by (\ref{bbJ}). In the topological twist background equation, 
(\ref{a4solns}) gives $\mafour=0$, so that $\langle \, \JRcur \rangle = -\mafive/\kappa_5^2$.
On the other hand, from (\ref{divJR}) we obtain the $U(1)_R$ anomaly equation
\beq
\diff *_4 \langle \,\JRcur \rangle \ = \ \frac{\ii}{8\kappa_5^2} F^I\wedge F^I~,\label{divJRFF}
\eeq
where $*_4$ denotes the Hodge duality operator on $(M_4,\met)$. Using equations (\ref{Ftwist}) and (\ref{outerJI})
this may be rewritten as
\beq
\diff *_4 \langle \,\JRcur \rangle \ = \ \frac{\ii}{32\kappa_5^2}(\mathcal{E}+\mathcal{P})\, \vol_4~, \label{divJREP}
\eeq
where $\mathcal{E}$ and $\mathcal{P}$ are the Euler and Pontryagin densities, (\ref{EandP}).
On a compact $M_4$ without boundary these integrate to
$\int_{M_4}\mathcal{E}\, \vol_4 = 32\pi^2\chi(M_4)$, $\int_{M_4}\mathcal{P}\, \vol_4 = 48\pi^2 \sigma(M_4)$, 
so that integrating (\ref{divJREP}) over $M_4$ gives\footnote{A little less laboriously we can instead note 
that $F^I$ is the curvature of the bundle of self-dual two-forms $\Lambda_2^+M_4$, and the integral of the right hand side of (\ref{divJRFF}) is proportional to 
the first Pontryagin class $p_1(\Lambda_2^+M_4) = 2\chi(M_4)+3\sigma(M_4)$.}
\beq
\int_{M_4}\diff *_4 \langle \,\JRcur \rangle \ = \ \frac{\ii\pi^2}{2\kappa_5^2}\left[2\chi(M_4)+3\sigma(M_4)\right]~.\label{intdivJR}
\eeq
It follows that if $\mafive$, or equivalently $\langle \, \JRcur\rangle$, is a \emph{global} one-form on $M_4$, then
by Stokes' theorem the left hand side of (\ref{intdivJR}) is zero, implying the topological constraint
\beq
2\chi(M_4)+3\sigma(M_4)\ = \ 0~.\label{topconstraint}
\eeq
Indeed, in section \ref{SecRomans} we noted that we are studying gravitational saddle points in
 the real Euclidean Romans theory, where the $U(1)_R$ gauge field $\cA$ is a (purely imaginary) global one-form. Related to this, the $U(1)_R$ symmetry effectively becomes an $SO(1,1)_R$ symmetry after Wick rotation, 
as also emphasized in \cite{Klare:2013dka} (see also \cite{Pestun:2007rz}). A number of gravity expressions that we shall obtain below 
only make sense if $\mafive$ is interpreted as a global one-form on $M_4$, at least in the set-up we have defined. 
Thus (\ref{topconstraint}) already restricts the topology of $M_4$. Interestingly, in section~\ref{SecTopAdSCFT} 
we shall see that (\ref{topconstraint}) also plays an important role  in the dual TQFT. Specifically, 
if (\ref{topconstraint}) does not hold, the partition function is zero!\footnote{In passing we note that (\ref{topconstraint}) corresponds (with an appropriate choice of orientation) to equality 
in the Hitchin-Thorpe inequality. In particular the only Einstein manifolds satisfying this condition are the flat torus, a K3 surface, 
or a quotient thereof \cite{hitchin}. A non-example is $S^4$, for which $2\chi(S^4)+3\sigma(S^4)=4$. On the other hand, for 
a complex surface (\ref{topconstraint}) is equivalent to $\int_{M_4}c_1\wedge c_1=0$, where $c_1=c_1(M_4)$ is the first Chern class of the holomorphic tangent bundle (the anti-canonical class).}

\subsection{Supersymmetric expansion}\label{SecSUSYexpand}

In this section we continue to expand the bulk spinor equations to higher order in $z$. From this we extract further information about some of the fields which are not fixed, in terms of boundary data, by the bosonic equations of motion. 
We will continue to use the boundary conditions appropriate to the topological twist. In particular we note that 
the boundary $B$-fields $b^\pm=0$ in this case, and that setting the bulk $\mathcal{B}^\pm=0$ is a consistent 
truncation of the Euclidean $\mathcal{N}=4^+$ theory. Moreover, in this case the bulk spinors $\epsilon^\pm$ satisfy decoupled 
equations, and since the leading order term $\los^-=0$ it is then also consistent to set the bulk $\epsilon^-=0$. 
We henceforth work in this truncated theory. 
This subsection is somewhat technical. All of the relevant formulas that we need 
in section \ref{SecVary} are in any case summarized in that section, and a reader uninterested in the details may safely 
skip the present subsection.

The frame, spin connection and spinor expansions beyond the leading order given in section \ref {SecBoundaryKSE} will be needed, so we first give details of these. The frame expansion is 
\beq
\label{eq:FrameExpansion}
\mathtt{e}^{\overline{i}}_i \ = \ \ex^{\overline{i}}_i + z^2( \ex^{(2)})^{\overline{i}}_i + z^4 \left[ (\log z)^2(\mathring{\ex}^{(4)})^{\overline{i}}_i + \log z (\tilde{\ex}^{(4)})^{\overline{i}}_i + (\ex^{(4)})^{\overline{i}}_i \right] + o(z^4) \, ,
\eeq
where in particular $\ex^{\overline{i}}_i $ is a frame for the boundary metric. The additional spin connection components we will need are
\begin{align}
\label{eq:SpinConnection5d}
	\Omega_i{}^{\overline{zi}} \ =& \ \frac{1}{z}\mathtt{e}^{\overline{i}}_i - \tfrac{1}{2} \mexp^{jk} \mathtt{e}^{\overline{i}}_j\partial_z\mexp_{ik}\, \qquad
	\Omega_z{}^{\overline{ij}} \ = \ \ \mexp^{ij}\mathtt{e}^{[\overline{i}}_i \partial_z\mathtt{e}^{\overline{j}]}_j \, .
\end{align}
The bulk spinor has $\epsilon^-=0$ in our truncated theory, and we thus henceforth drop the superscript 
on $\epsilon^+\rightarrow\epsilon$, $\los^+\rightarrow\los$ (we hope this abuse of notation won't lead to any confusion).
The bulk spinor then has the following  expansion
\begin{equation}
	\epsilon \ =  \ z^{-1/2} \los + z^{3/2}\varepsilon^3 + z^{5/2} ( \log z \, \tilde{\varepsilon}^5 + \varepsilon^5 ) + z^{7/2} \big( (\log z)^2 \, \mathring{\varepsilon}^7 + \log z \, \tilde{\varepsilon}^7 + \varepsilon^7 \big) + o(z^{7/2})\, ,\label{genspinorexp}
\end{equation}
where $\los$ is constant with positive chirality under $-\gamma_{\bar{z}}$. As in equation (\ref{losreal})
the bulk spinor $\epsilon$ satisfies the reality condition
\beq
\epsilon^c \ \equiv \ \ii \sigma_2  \mathscr{C}\epsilon^* \ = \ \epsilon~.
\eeq

We start by analysing the bulk dilatino equation. At lowest order we find
\begin{equation}
	0 \ = \ \Xb\, \los + \tfrac{\ii}{6}F^I\cdot(\sigma^I \los) \ = \ \left( \Xb + \tfrac{1}{12} R \right) \los \, ,
\end{equation}
which is satisfied identically, where we have used (\ref{X2twist}) and (\ref{Ftwist}). At the next order we find
\begin{equation}
	\ii \aIfour\cdot\left(\sigma_I \los \right) = - \tfrac{1}{4}(\dd R)\cdot \los \, . \label{aI4matrix}
	\end{equation}	
This is effectively a matrix equation, of which we shall see many more. Components of such equations 
may be extracted by first noting  that
\beq
\los \ = \ \begin{pmatrix}\chi \\ -\mathscr{C}\chi^*\end{pmatrix}~,
\eeq
in the notation of section \ref{SecTT}. For example, one can then take the first component of (\ref{aI4matrix}), 
and apply $\bar\chi\gamma_j$ on the left. Taking the real part, and 
using the definitions (\ref{JIbilinears}) of $\Jb^I$ in terms of spinor bilinears, one obtains
\begin{equation}\label{eq:FirstOrderDilatino}
	(\aIfour)^i \, \Jb^{I}_{ij} \ = \ \tfrac{1}{4} \nabla_j R \, . 
\end{equation}
We shall make use of similar manipulations throughout this subsection. Focusing on~(\ref{eq:FirstOrderDilatino}), 
recall that $\aIfour$ is already fixed in terms of the $SU(2)$ covariant divergence of $F^I$, via equation (\ref{a4solns}). 
The latter reads $(\aIfour)_i=\frac{1}{2}\mathcal{D}^j F^{I}_{ij}$. Starting from this and~(\ref{Ftwist}), and using the identity $\alpha_{pq} \Jb^I_m{}^p \Jb^I_n{}^q = \alpha_{mn} - 2 (*\alpha)_{mn}$, where $\alpha_{pq}$ is any two-form, one can show that (\ref{eq:FirstOrderDilatino}) 
is an identity. We may then differentiate \eqref{eq:FirstOrderDilatino} and, upon using the quaternionic K\"ahler equation \eqref{QKeqn}, we obtain
\begin{equation}
	(\mathcal{D}\aIfour)^{ij} \Jb^{I}_{ij} \ =\  - \tfrac{1}{2}  \nabla^2 R \, . \label{DaI4}
\end{equation}
This relation appears frequently hereafter.

At the next order in the dilatino equation we find an equation involving several undetermined fields:
\begin{equation}
	\ii \aIfive\cdot\left(\sigma_I \los\right) \ = \  \left( {2} \ii \mafive + {3}\dd \Xs + \tfrac{1}{8}\dd R\right)\cdot \los \, , 
\end{equation}	
from which we similarly extract
\begin{equation}
	(\aIfive)^i \Jb^{I}_{ij} \ = \ - {2}\ii (\mafive)_j -{3}\nabla_j \Xs - \tfrac{1}{8}\nabla_j R~.\label{eq:aI5}
\end{equation}
From this expression, taking a covariant derivative and symmetrizing indices gives
\begin{align}
3\nabla_i\nabla_j \Xs \ =& \ \mathcal{D}_{(i}(\aIfive)^{k} {\Jb^I_{j)k}} - 2\ii \nabla_{(i} (\mafive)_{j)} -\tfrac{1}{8}\nabla_i\nabla_j R~. \label{nnX1}
\end{align}
At higher order still we have
\begin{equation}
	\Xseven\, \los \ = \ \Xb (1+\gamma_{\overline{z}}) \varepsilon^3 - \tfrac{\ii}{12} \mathcal{D}\aIfour \cdot \left(\sigma_I\los\right)\, .
\end{equation}	
As $\los$ has positive chirality we can act with $P_- = \frac{1}{2} ( 1 + \gamma_{\bar{z}} )$ 	to deduce that $\varepsilon^3$ also has positive chirality. It then follows that
\begin{equation}
	\Xseven \ = \ -\tfrac{1}{24} (\mathcal{D}\aIfour)^{ij} \Jb^{I}_{ij} \ = \ \tfrac{1}{48}  \nabla^2 R~. 
\end{equation}
 where we have used \eqref{DaI4}. This expression for $\Xseven$ is equivalent to that in \eqref{nablaXb}, for the topological twist. 
	Finally, at order $\mathcal{O}(z^{7/2})$ we have
\begin{align}
	\Xeight \, \los \ =& \ - \tfrac{1}{2} \Xseven \, \los - \tfrac{1}{2} \Xb \, \varepsilon^3 -\tfrac{\ii}{12}\Big[ ( \mathcal{D}\aIfive ) \cdot \left( \sigma_I\los \right) - {2}\mffive\cdot \los + F^I\cdot (\sigma_I\varepsilon^{3}) \Big]\nn\\
	& \ - \tfrac{\ii}{12}\, \ex^i_{\overline{i}}\, (\ex^{(2)})^j_{\overline{j}}F^I_{ij}\gamma^{\overline{ij}}(\sigma_I \los) \, . \label{X8interm}
\end{align}	
Here $\te^i_{\overline{i}}$ is the inverse frame to $\te^{\overline{i}}_i$, with $\ex^i_{\overline{i}}$ and $(\ex^{(2)})^i_{\overline{i}}$ 
being coefficients in its expansion, precisely as in (\ref{eq:FrameExpansion}). We have also defined $\mffive=\diff\mafive$.
Since $\varepsilon^3$ is so far undetermined, we cannot yet extract an expression for $\Xeight$. This concludes the expansion of the bulk dilatino equation. 

Turning next to the bulk gravitino equation,
at lowest order in the $z$ direction we find, after using the fact that $\varepsilon^3$ has positive chirality, that
\beq\label{eq:Epsilon3}
\varepsilon^3 \ = \ \tfrac{1}{48}R\, \los - \tfrac{1}{4}g^{ij}\, \ex^{\overline{i}}_i \, (\ex^{(2)})^{\overline{j}}_j \gamma_{\overline{ij}}\, \los\, .
\eeq
As a metric defines the frame only up to an arbitrary local $SO(4)$ rotation, it is convenient to gauge fix this 
arbitrariness. A consistent gauge choice is $( \ex^{(2)})^{\bar{i}}_i = \frac{1}{2} ( \mexp^2 )^{\bar{i}}{}_{\bar{j}} \, \ex^{\bar{j}}_i$ and $(\ex^{(2)})_{\bar{i}}^i = -\frac{1}{2}\ex_{\bar{j}}^i \, ( \mexp^2 )^{\bar{j}}{}_{\bar{i}}$, where recall that $\mexp^2$ is fixed in 
terms of the boundary Schouten tensor via (\ref{g2}). 
This then implies that
\begin{equation}
	g_{ij}\, \ex^i_{\overline{i}}\, (\ex^{(2)})^j_{\overline{j}} \ = \  - \tfrac{1}{2}\mexp^2_{\overline{ij}}\, , \qquad  g^{ij}\, \ex^{\overline{i}}_i\,  (\ex^{(2)})^{\overline{j}}_j \ =\  \tfrac{1}{2} ( \mexp^2 )^{\overline{ij}}\, ,
\end{equation}
and, being symmetric, their contraction with any anti-symmetric tensor automatically vanishes. Consequently, this gauge choice reduces the relation between the spinors $\los$ and $\varepsilon^3$ to simply
\begin{equation}
	\varepsilon^3 \ = \ \tfrac{1}{48}R \, \los \, .\label{epsilon3}
\end{equation}
Having found this relation we may substitute for $\varepsilon^3$ into the right hand side of \eqref{X8interm}, extract $\Xeight$ and then substitute for $\mexp^2$, $\Xb$, $\Xseven$ and $F^I$ to obtain
\begin{align}
	\Xeight \ =& \ \tfrac{1}{288} R^2 - \tfrac{1}{48} R_{kl}R^{kl} - \tfrac{1}{96}\nabla^2 R - \tfrac{1}{24} \left( \mathcal{D} \aIfive \right)^{ij} \Jb^{I}_{ij} \, . \label{X8final}
\end{align}
Here strictly speaking we have taken the real part of this equation, where the term involving $\mffive$ is purely imaginary, 
and thus doesn't appear.
Using the trace of (\ref{nnX1}), together with several other equations derived so far,  one can check that the expression (\ref{X8final}) for $\Xeight$ agrees with 
the expression  (\ref{nablaXs}), obtained from the equations of motion.

At the next orders we find
\begin{align}
	(5-\gamma_{\overline{z}})\, \varepsilon^5 \ =& \ - 2\, \tilde{\varepsilon}^5 +  2(\ii\mafive + \dd \Xs )\cdot \los\, , \label{eq:epsilon5Raw}\\
	(5-\gamma_{\overline{z}})\, \tilde{\varepsilon}^5 \ =& \ \tfrac{2\ii}{3} \aIfour\cdot\left(\sigma_I \los\right) = - \tfrac{1}{6}\dd R\cdot \los\, .\label{eq:Weird}
\end{align}
We could continue and analyse higher order terms in this $z$ component of the 
gravitino equation, but the subsequent expressions are not required, nor particularly enlightening, and so we stop here.

The remaining equation to study is the $i$ direction of the gravitino equation. Crucially this involves the spin connection components $\Omega_i{}^{\overline{zi}}$, which introduce the metric expansion fields from \eqref{metricexp}. Of course, the leading order equation is satisfied by construction. Remarkably, at the next order we find a non-trivial equation which is also identically satisfied given the chirality of $\varepsilon^3$ and the algebraic properties of the Riemann tensor. At the following order we find another condition on $\tilde{\varepsilon}^5$: 
\beq
  \gamma_{\overline{i}}\left[ {3\ii}(1+\gamma_{\overline{z}})\tilde{\varepsilon}^5 + \aIfour\cdot (\sigma_I \los)\right] \ = \ 0\, ,
\eeq
which, used in conjunction with \eqref{eq:Weird}, allows us to determine
\begin{equation}
\label{eq:EpsilonTilde5}
\gamma_{\overline{z}}\tilde{\varepsilon}^5 \ = \ \tilde{\varepsilon}^5 \, , \qquad \tilde{\varepsilon}^5 \ = \ - \tfrac{1}{24}\dd R \cdot \los \, .
\end{equation}
We now substitute $\tilde{\varepsilon}^5$ into equation \eqref{eq:epsilon5Raw}:
\beq
( 5 - \gamma_{\overline{z}} ) \varepsilon^5 \ = \ \left( 2\ii \mafive + 2\dd \Xs + \tfrac{1}{12}\dd R \right)\cdot \los\, .
\eeq
Acting on this last equation with $\gamma_{\bar{z}}$, and taking the difference, implies that $\varepsilon^5$ is a negative chirality spinor: $\gamma_{\bar{z}} \varepsilon^5 = \varepsilon^5$. We thus find 
\beq
\varepsilon^5 \ = \ \left( \tfrac{\ii}{2} \mafive + \tfrac{1}{2}\dd \Xs + \tfrac{1}{48}\dd R \right)\cdot \los\, . \label{eq:Epsilon5}
\eeq

At the next order we begin to see the metric fields appearing:
\beq
h^0_{\overline{ij}}\gamma^{\overline{j}}\los \ = \ -\tfrac{1}{288}R^2\gamma_{\overline{i}}\los - \tfrac{1}{2}\gamma_{\overline{i}}(1+\gamma_{\overline{z}})\mathring{\los}^7\, .
\eeq	
Using the chiral projector $P_-$ again we see that $\mathring{\varepsilon}^7$ has positive chirality, and we may extract $h^0$:
\begin{equation}
	h^0_{ij}  \ = \  - \tfrac{1}{288} R^2 g_{ij} \, .
\end{equation}
This agrees with the expression $h^0_{ij} = - \frac{1}{2} g_{ij} \Xb^2$, given by equation (\ref{h0eqn}), 
derived from the  expansion of the bosonic field equations.
The next order gives
\begin{align}
	h^1_{\overline{ij}}\gamma^{\overline{j}} \los \ =& \ - \tfrac{1}{2} \gamma_{\bar{i}} ( 1 + \gamma_{\bar{z}} ) \, \tilde{\los}^7 - \tfrac{1}{2} h^0_{\overline{ij}} \gamma^{\bar{j}} \los - \Xb \Xs \gamma_{\overline{i}}\, \los + \nabla_{\overline{i}}\tilde{\los}^5 + \tfrac{\ii}{{2}}A^I_{\overline{i}}(\sigma_I\tilde{\los}^5)\nn \\
	&\ - \tfrac{\ii}{24} \Xb (\gamma_{\overline{i}}{}^{\overline{jk}}-4\delta^{\overline{j}}_{\overline{i}}\gamma^{\overline{k}}) F^I_{\overline{jk}}(\sigma_I\los ) + \tfrac{\ii}{24}(\gamma_{\overline{i}}{}^{\overline{jk}}-4\delta^{\overline{j}}_{\overline{i}}\gamma^{\overline{k}})(\mc{D}\aIfour)_{\overline{jk}}(\sigma_I\los )\, . \label{h1raw}
\end{align}
 As before, we can show that $\tilde{\los}^7$ has positive chirality and hence drops out of \eqref{h1raw}. Now using the definition of $\tilde{\los}^5$ in \eqref{eq:EpsilonTilde5} allows us to write everything acting on the spinor $\los$. After using the intermediate result
\begin{equation}
	- \tfrac{1}{4} {\Jb^I{}_{(i}{}^k} ( \mathcal{D}\aIfour )_{j)k} \ = \ - \tfrac{1}{8} \left( R_i{}^k R_{jk} + R_{iklj} R^{kl} - \nabla^2 R_{ij} + \tfrac{1}{2} \epsilon_{(j|kmn|} R^{kl} R^{mn}{}_{i)l} \right) \, ,
\end{equation}
and substituting for the known expressions, we can then read off $h^1_{ij}$:
\begin{align}
	h^1_{ij} \ =& \ \tfrac{1}{192} g_{ij} R^2 + \tfrac{1}{12} g_{ij} R \Xs - \tfrac{1}{48}R R_{ij} - \tfrac{1}{24} \nabla_i\nabla_j R - \tfrac{1}{48} g_{ij} \nabla^2 R \nn \\
	&- \tfrac{1}{8} \left( R_i{}^k R_{jk} + R_{iklj} R^{kl} - \nabla^2 R_{ij} + \tfrac{1}{2} \epsilon_{(j|kmn|} R^{kl} R^{mn}{}_{i)l} \right) \, . \label{h1expr2}
\end{align}
	Once again, we have found another expression for something we have already derived: $h^1_{ij}$ is also given by 
	equation (\ref{h1eqn}). However, in this instance the equality of the two expressions \eqref{h1expr2} and (\ref{h1eqn})
is non-trivial. It is equivalent to the equation
\begin{align}
	0 \, =& \   (  R R_{ij} - 2 R_i{}^k R_{jk} + 2 R_{iklj} R^{kl} + R_{mnik} R^{mn}{}_j{}^k ) - \tfrac{1}{4} g_{ij} ( R^2 - 4 R_{kl} R^{kl}+ R_{mnkl} R^{mnkl} ) \nn \\
	& \ + \tfrac{1}{2} \big[ \epsilon_{mnpq} \big( - \tfrac{1}{4} g_{ij} R^{mn}{}_{kl} R^{pqkl} +g_{jk} R^{mn}{}_{il} R^{pqkl} \big) - 2 \epsilon_{(j|kmn|} R^{kl} R^{mn}{}_{i)l} \big] \, . \label{Berger}
\end{align}	
The first line quite remarkably is known to be zero for any Riemannian four-manifold, 
and is called Berger's identity \cite{Berger}. One can also show that the second line is equal to zero, which amounts to an algebraic 
identity that holds  for any tensor sharing the algebraic symmetries of the Riemann tensor.

Finally, at the last order we find\footnote{Of course, knowing $h^1_{\overline{ij}}$ we could write an expression for $\mexp^4_{\overline{ij}}$ alone, but it is only the combination $4 \mexp^4_{\overline{ij}} + h^1_{\overline{ij}}$ which we shall need in the next section.}
\begin{align}
	( 4 \mexp^4_{\overline{ij}} + h^1_{\overline{ij}} ) \gamma^{\overline{j}}\los \ = & \ - 2 \gamma_{\bar{i}} ( 1 + \gamma_{\bar{z}} ) \los^7 + 4 \left(\nabla_{\overline{i}}\, \varepsilon^5 + \tfrac{\ii}{{2}}A^I_{\overline{i}}(\sigma_I\varepsilon^5)\right) - 2 \Xs^2 \gamma_{\overline{i}}\, \los  - 2 \mexp^2_{\overline{ij}} \gamma^{\bar{j}} \los^3  \nn \\
	&\ +  \tfrac{\ii}{6}(\gamma_{\overline{i}}^{\ph{i}\overline{jk}}-4\delta_{\overline{i}}^{[\overline{j}} \gamma^{\overline{k}]})\Big[ (\mc{D}\aIfive)_{\overline{jk}}(\sigma_I\los )+ (\mffive)_{\overline{jk}}\los + F^I_{\overline{jk}}(\sigma_I\varepsilon^{3}) - \Xs F^I_{\overline{jk}}(\sigma_I\los)\nn\\
	&\ + 2 \ex^j_{\overline{j}} (\ex^{(2)})^k_{\overline{k}} F^I_{jk} (\sigma_I\los)\Big]   - 2 \Big[ \ex^i_{\overline{i}} \, (\ex^{(2)})^j_{\overline{j}} +(\ex^{(2)})^i_{\overline{i}} \, \ex^j_{\overline{j}} \Big] \mexp^2_{ij} \gamma^{\bar{j}} \los\, . 
\end{align}
Again there is a positive chirality condition on $\los^7$ which removes it from the above equation. Using the many intermediate results we have derived, we then find
\begin{align}
	4 \mexp^4_{ij} + h^1_{ij} \ =& \ \, 2\nabla_i\nabla_j\left( \Xs + \tfrac{1}{24}R\right) + 2\ii\nabla_{(i}(\mafive)_{j)} + \left( \Xs - \tfrac{1}{12} R\right)R_{ij} \nn \\
	&+ g_{ij} \left( -\tfrac{1}{6} R \Xs - 2 \Xs^2 + \tfrac{1}{12}R_{kl}R^{kl} \right) + \tfrac{1}{4}R_{ik}R^{k}{}_{j} \nn \\
	&- \tfrac{1}{8} \epsilon^{mnk}{}_{j}R_{mnli}R_{k}{}^{l} + \tfrac{1}{4} R_{iklj}R^{kl}+ \tfrac{1}{3} [2\mathcal{D}\aIfive - *(\mathcal{D}\aIfive) ]_{(i|k|} {\Jb^{Ik}}_{|j)} \, .\label{g4h1}
\end{align}
	

\section{Metric independence}\label{SecVary}

Our aim in this section is to show that, for any supersymmetric asymptotically locally hyperbolic solution 
to the Euclidean $\mathcal{N}=4^+$ supergravity theory, with the topologically twisted boundary conditions on 
an arbitrary Riemannian four-manifold $(M_4,g)$, 
the variation (\ref{holWard})  of the holographically renormalized action is identically zero.  
As explained in the introduction, this implies that the right hand side of (\ref{saddle}) 
is independent of the choice of metric $g$, precisely as expected for the holographic dual of a topological QFT. 
We find that this is indeed the case, using the minimal holographic renormalization scheme 
described in section \ref{SecHoloRenormalization}. We comment further on this 
at the end of section \ref{SecProof}.

\subsection{Variation of the action}

As discussed in section \ref{SecTT}, the Donaldson-Witten topological twist corresponds to the following boundary conditions on the supergravity fields on $M_4$:
\begin{align}
	0 \ = \ b^\pm \ = \ \ma \ = \ \los^- \, , \qquad \Xb \ = \ - \tfrac{1}{12}R \, , \qquad A^I \ = \ \tfrac{1}{2}{\omega_i{}^{\overline{jk}}} {\Jb^I_{\overline{jk}}} \, \dd x^i \, . \label{TTwistBulk}
\end{align}
Here the boundary Riemannian metric $g_{ij}$ on $M_4$ is arbitrary, with $\omega_i^{\ \overline{jk}}$ being 
the spin connection, $R$ being the Ricci scalar curvature, and the triplet of self-dual two-forms $\Jb^I$ being given by 
(\ref{JIframe}). The holographic Ward identity for the variation of the renormalized action (\ref{Actionfinal}) with 
respect to general variations of the non-zero boundary fields is
\begin{align}
	\delta S \, = \, \delta_g S + \delta_{A^I}S + \delta_{\Xb}S 
	\, = \,	\int_{\partial \B_5=M_4} \dd^4 x \, \sqrt{\det g}\, \Big[ \tfrac{1}{2} T_{ij} \delta g^{ij}  + \mathscr{J}^{i}_I \delta A^I_i+ \Xi\,  \delta \Xb \Big] \, .\label{holWardagain}
\end{align}
It is worth pausing to consider carefully why this equation holds. A variation of the boundary data on $M_4$ will 
induce a corresponding variation of the bulk solution that fills it. However, we are evaluating 
the action on a solution to the equations of motion, and by definition these are stationary points of the 
 bulk action. Thus the resulting variation of the on-shell action is necessarily a boundary term, 
and this is the expression on the right hand side of (\ref{holWardagain}). This argument requires that the equations of motion 
are solved everywhere in the interior of $\B_5$: if the latter has internal boundaries, or singularities, 
the above in general breaks down, and one will encounter additional terms around these 
boundaries/singularities on the right hand side of (\ref{holWardagain}). 

For the 
 topological twist all boundary fields are determined by the metric $g_{ij}$. Since $\Xb=-\frac{1}{12}R$, to compute 
 $\delta \Xb$ we need the 
 variation  of the Ricci scalar:
\beq
\delta R \ = \ R_{ij}\delta g^{ij} + \nabla_i\left( g^{jk}\delta \Gamma^i_{jk} - g^{ij}\delta \Gamma^k_{jk}\right),
\eeq
with the variation of the Christoffel symbols being
\beq
\delta \Gamma^i_{jk} \ = \ \tfrac{1}{2} g^{il} \left(\nabla_k\delta g_{lj} + \nabla_j\delta g_{lk} - \nabla_l\delta g_{jk}\right)\, . \label{varyChrist}
\eeq
After integrating by parts twice we obtain
\begin{align}
\delta_{\Xb} S \ = \ - \frac{1}{12} \int_{\partial \B_5} \bigg[\big( &\, \Xi R_{ij} +g_{ij}\nabla^2\Xi - \nabla_i\nabla_j\Xi\, \big)\delta g^{ij}\, \vol_4 + \frac{1}{\kappa_5^2}\mathscr{D}_{\Xb} \vol_4 \,\bigg] ,
\end{align}
where $\vol_4\equiv \sqrt{\det g}\, \diff^4x$ is the Riemannian volume form on $(M_4,g)$, and all geometric quantities appearing are computed using the boundary metric $g_{ij}$.
Substituting the value of $\Xi$ from \eqref{Xi} leads to
\beq
\label{eq:VariationX}
\delta_{\Xb}S \ = \  - \frac{1}{4\kappa^2_5} \int_{\partial \B_5} \big[\left( \Xs R_{ij} + g_{ij}\nabla^2 \Xs - \nabla_i\nabla_j \Xs \right)\delta g^{ij}\, \vol_4 + \tfrac{1}{3} \mathscr{D}_{\Xb} \vol_4\,\big] ,
\eeq
where the total derivative term is
\begin{equation}
	\mathscr{D}_{\Xb} \ \equiv \ - 3 \nabla_i \Big[ \nabla^k \Xs g^{ij} \delta g_{jk} - \nabla^i \Xs g^{jk} \delta  g_{jk} - \Xs g^{jk} g^{il} (\nabla_k\delta g_{lj} - \nabla_l\delta g_{jk}) \Big]\, .
\end{equation}

For $\delta A^I_i$ we first need the variation of the spin connection. After a short calculation we have 
\begin{align}
	\delta \omega_i{}^{\overline{jk}} \ = \ \tfrac{1}{2} \, \ex^{l\overline{j}}\, \ex^{m\overline{k}} \left(\nabla_m \delta g_{il} - \nabla_l \delta g_{im}\right)\, .
\end{align}
Thus
\begin{align}
	\delta A^I_i \ = \ \tfrac{1}{2} \delta \omega_i{}^{\overline{jk}} \Jb^I_{\overline{jk}} \ = \ \tfrac{1}{2} ( \nabla_k \delta \mexp_{ij} ) \Jb^{I jk} \, . 
\end{align}
After integrating by parts, the $SU(2)_R$ current contribution is hence
\begin{align}
	\delta_{A^I} S \ =& \ -  \frac{1}{8\kappa^2_5} \int_{\partial \B_5} \bigg\{\left[ \mathcal{D}^k ( \aIfour + 2 \aIfive  )_i \, {\Jb^I_{jk}} \right] \delta g^{ij} \, \vol_4 + \mathscr{D}_{A^I} \vol_4 \,\bigg\} , \label{eq:VariationaI}
\end{align}
where we have substituted for the $SU(2)_R$ current using \eqref{calJi}, and used the quaternionic K\"ahler identity \eqref{QKeqn}. The object in square brackets is a tensor with indices $ij$: only the symmetric part contributes. The total derivative term is
\begin{align}
	\mathscr{D}_{A^I} \equiv \  \nabla_i \Big[ (\aIfour + 2\aIfive)^k {\Jb^{I ij}} \delta g_{jk} \Big] \, .
\end{align}

It remains to evaluate the stress-energy tensor contribution (\ref{Tij}) and combine it with \eqref{eq:VariationX} and \eqref{eq:VariationaI}. Doing so leads to
\begin{align}
	\delta S \ = \  \frac{1}{4\kappa_5^2}\int_{\partial \B_5}\left( \mathcal{T}_{ij} \, \delta g^{ij} \, \vol_4 + \mathscr{D}_S\, \vol_4 \right)\, , 
\end{align}
where the total derivative term is
\beq
\mathscr{D}_S \ \equiv \ -\tfrac{1}{3}\mathscr{D}_{\Xb}-\tfrac{1}{2}\mathscr{D}_{A^I}~,
\eeq
and 
\begin{align}
	\mathcal{T}_{ij} \ = \ \big[ &4 \mexp^4_{ij} + h^1_{ij} - 4 g_{ij}\big( t^{(4)} - \tfrac{1}{2} t^{(2,2)} - \tfrac{1}{8} u^{(1)} \big)  - 2 \mexp_{ij}^{2} t^{(2)} - 6 g_{ij} \Xs^2 \nn \\
	&+ \tfrac{1}{2} \big( \nabla^k \nabla_i \mexp^2_{jk} + \nabla^k \nabla_j \mexp^2_{ik} - \nabla^2 \mexp^2_{ij} - \nabla_i \nabla_j t^{(2)} \big) - \tfrac{1}{2} \mexp_{ij}^{2} R + \tfrac{1}{2} g_{ij} \big( \mexp^2_{kl} R^{kl} \big) \big] \nn \\
	&- \left( \Xs R_{ij} + g_{ij}\nabla^2 \Xs - \nabla_i\nabla_j \Xs \right) - \tfrac{1}{2} \left[ \mathcal{D}^k ( \aIfour + 2 \aIfive  )_{(i} \, {\Jb^I}_{j)k} \right] \, . \label{ZeroIntegrand}
\end{align}
Here the first two lines come from the stress-energy tensor (\ref{Tij}), while the last line combines (\ref{eq:VariationX}) and 
(\ref{eq:VariationaI}). Provided $M_4$ is a closed manifold, without boundary, the integral of the total derivative term is 
zero, and we have simply
\beq
\delta S \ = \  \frac{1}{4\kappa_5^2}\int_{\partial \B_5=M_4} \mathcal{T}_{ij} \, \delta g^{ij} \, \vol_4 ~.
\eeq
The tensor $\mathcal{T}_{ij}$ is thus an \emph{effective} stress-energy tensor, for variations of the 
renormalized on-shell action with respect to the boundary metric, all boundary data being determined by this choice of metric. 
Our claim that the on-shell action is invariant under an arbitrary metric deformation $\delta g_{ij}$ is thus equivalent to the statement that $\mathcal{T}_{ij} \equiv 0$, for every Riemannian four-manifold. Remarkably, despite there being several undetermined quantities in \eqref{ZeroIntegrand}, using the results of sections \ref{SecHoloRenormalization} and \ref{SecSUSYexpand} we will show that indeed  $\mathcal{T}_{ij}\equiv 0$ in the next subsection. 

\subsection{Proof that \texorpdfstring{$\delta S/\delta g_{ij}=0$}{dS/ddgij}}\label{SecProof}

We begin by substituting expressions from section \ref{SecFG} into (\ref{ZeroIntegrand}), which recall follow from the Fefferman-Graham expansion 
of the bosonic equations of motion. In particular we substitute for $\nabla^2 \Xs$ using equation (\ref{nablaXs}), as well as various  
metric quantities, except for the combination $4 \mexp^4_{ij} + h^1_{ij}$. With the topological twist boundary conditions~\eqref{TTwistBulk} this leads to the expression
\begin{align}
	\mathcal{T}_{ij} \ =& \ \left( \tfrac{1}{12} R - \Xs \right)R_{ij} - \tfrac{1}{2} R_{ik} R^k{}_j - \tfrac{1}{2} R_{iklj} R^{kl} - \tfrac{1}{4} \nabla_i \nabla_j R + \nabla_i\nabla_j\left( \Xs + \tfrac{1}{6} R \right) \nn \\
	&+ \tfrac{1}{4}\nabla^2R_{ij} + g_{ij}\left(2 \Xs^2 - \tfrac{1}{72} R^2 + \tfrac{1}{6} R \Xs - \tfrac{1}{24}\nabla^2R + 4\Xseven + 4\Xeight\right) \nn \\
	&+ 4\mexp^4_{ij} + h^1_{ij} - \tfrac{1}{{2}}\left[ \mathcal{D}^k ( \aIfour + 2 \aIfive  )_{(i} \, {\Jb^I}_{j)k} \right] 	\, .\label{Tijagain}
\end{align}
 In particular we have used the identity
\beq\label{ddcurv}
 - \tfrac{1}{2} \nabla_k \nabla_{(i} R^k{}_{j)} \ = \  - \tfrac{1}{2} R_{ik} R^k{}_j - \tfrac{1}{2} R_{iklj} R^{kl} - \tfrac{1}{4} \nabla_i \nabla_j R~,
 \eeq
 in 
deriving (\ref{Tijagain}).

The equations of motion, or equivalently supersymmetry conditions, determine
\begin{align}
	\Xseven \ =& \ \tfrac{1}{48} \nabla^2 R \, , \quad \Xeight \ = \ \tfrac{1}{288} R^2 - \tfrac{1}{48} R_{kl}R^{kl} - \tfrac{1}{96}\nabla^2 R - \tfrac{1}{12} \left( \mathcal{D} \aIfive \right)^{ij} \Jb^{I}_{ij} \, .
\end{align}
On the other hand, in section \ref{SecSUSYexpand} the expansion of the supersymmetry conditions led to the expression (\ref{g4h1}), which 
we repeat here:
\begin{align}
	4 \mexp^4_{ij} + h^1_{ij} \ =& \ \, 2\nabla_i\nabla_j\left( \Xs + \tfrac{1}{24}R\right) + 2\ii\nabla_{(i}(\mafive)_{j)} + \left( \Xs - \tfrac{1}{12} R\right)R_{ij} \nn \\
	&+ g_{ij} \left( -\tfrac{1}{6} R \Xs - 2 \Xs^2 + \tfrac{1}{12}R_{kl}R^{kl} \right) + \tfrac{1}{4}R_{ik}R^{k}{}_{j} \nn \\
	&- \tfrac{1}{8} \epsilon^{mnk}{}_{j}R_{mnli}R_{k}{}^{l} + \tfrac{1}{4} R_{iklj}R^{kl}+ \tfrac{1}{3} [ 2\mathcal{D}\aIfive - *(\mathcal{D}\aIfive) ]_{(i|k|} {\Jb^{Ik}}_{|j)} \, .
\end{align}
Substituting into (\ref{Tijagain}), after several immediate cancellations we are left with
\begin{align}
	\mathcal{T}_{ij} & \, =  \tfrac{1}{4} \nabla^2 R_{ij}- \tfrac{1}{8} \epsilon^{mnk}{}_{j}R_{mnpi}R_{k}{}^{p} - \tfrac{1}{4} R_{ik} R^k{}_j - \tfrac{1}{4} R_{iklj} R^{kl} + 3 \nabla_i\nabla_j \Xs  -\tfrac{1}{{2}} \mathcal{D}^k ( \aIfour )_{(i} {\Jb^I}_{j)k} \nn \\
	& + 2\ii\nabla_{(i}(\mafive)_{j)}  - \tfrac{1}{6} g_{ij} \left( \mathcal{D} \aIfive \right)^{kl} J^{I}_{kl} + \tfrac{1}{3} ( 2\mathcal{D}\aIfive - *\mathcal{D}\aIfive )_{(i|k|} \Jb^{I k}{}_{j)} -  \mathcal{D}^k ( \aIfive )_{(i} {\Jb^I}_{j)k}~.\label{Tijagainagain}
\end{align}
Using the expression 
\beq
(\aIfour)_i \ = \ -\tfrac{1}{4}\Jb^I_{mn} \nabla_j R^{mnj}_{\ \ \ \ \,  i}~,
\eeq
together with  the contracted second Bianchi identity, we find that
\begin{align}
	\mathcal{D}^k ( \aIfour )_{i} \Jb^I_{jk} \ = \ - \tfrac{1}{2} \epsilon_{j}{}^{kmn} \nabla_k\nabla_m R_{ni} - \tfrac{1}{2}\nabla^k \nabla^l R_{jkli}\, .
\end{align}
Substituting this expression, together with equation (\ref{nnX1}),  into $\mathcal{T}_{ij}$ in (\ref{Tijagainagain}), we arrive at
\begin{align}
	\mathcal{T}_{ij} \ =& \ \ \ \tfrac{1}{4} \nabla^2 R_{ij} -\tfrac{1}{8}\nabla_i\nabla_j R+ \tfrac{1}{4}\nabla^k \nabla^l R_{jkli} - \tfrac{1}{4} R_{ik} R^k{}_j - \tfrac{1}{4} R_{iklj} R^{kl} \nn \\
	& - \tfrac{1}{6} g_{ij} \left( \mathcal{D} \aIfive \right)^{kl} J^{I}_{kl} + \tfrac{1}{3} [ 2\mathcal{D}\aIfive - *(\mathcal{D}\aIfive) ]_{(i|k|} \Jb^{I k}{}_{j)} - ( \mathcal{D}\aIfive )_{(i|k|} \Jb^{I k}{}_{j)} \nn \\
	&+ \tfrac{1}{8}\epsilon_{j}{}^{kmn} ( 2 \nabla_k\nabla_m R_{ni} - R_{mni}{}^l R_{kl} )\, \nn\\
	=& \ 0~.
\end{align}
Here, remarkably, each 
 of  the three lines vanishes separately. The first line is zero using again (\ref{ddcurv}) and the contracted second Bianchi identity, whilst the terms in the second line combine 
to give zero 
 after using the self-duality property of the $\Jb^I$ tensors to remove the Hodge dual acting on the field strength $\mathcal{D}\aIfive$. The final line is zero after applying the Ricci identity for a rank two covariant tensor, followed by the first Bianchi identity and using the symmetry of the summed indices. 
 
We emphasize again that this proof that $\delta S/\delta g_{ij}=0$ uses the minimal holographic renormalization scheme defined in section 
\ref{SecHoloRenormalization}. Up to finite counterterms in (\ref{Ifinite}) that are topological invariants, which 
 have identically zero variations, another choice of scheme would spoil the above result. 
Another important comment is that the original path integral arguments in \cite{Witten:1988ze} 
are essentially classical (see footnote 10 of \cite{Witten:1988ze}). In particular there might 
have been an anomaly, implying that the partition function (and other correlation functions) 
are not invariant under arbitrary metric deformations. In this case, the topological twist 
would not have led to a TQFT. This might seem like a strange comment, given that the topologically 
twisted $\mathcal{N}=2$ Yang-Mills theory of \cite{Witten:1988ze} at least formally reproduces Donaldson theory, which of course 
certainly does rigorously define diffeomorphism invariants of $M_4$. However,
 it has recently been 
argued that precisely such an anomaly exists for four-dimensional rigid $\mathcal{N}=1$
supersymmetry \cite{Papadimitriou:2017kzw, An:2017ihs}. The computations in these papers 
are in fact holographic, and rely on the fact that in AdS/CFT the semi-classical gravity computation 
is a fully quantum computation on the QFT side, including any potential anomalies. Specifically, 
it is argued that there is an anomalous transformation of the supercurrent under rigid supersymmetry 
on the conformal boundary, 
implying that the partition function is not invariant under certain metric deformations that 
are classically $\mathcal{Q}$-exact. 
These particular anomalous transformations were first 
discovered  in \cite{Genolini:2016sxe, Genolini:2016ecx}, via essentially the same computation 
we have followed in this paper, although this was not interpreted
as an anomaly in  \cite{Genolini:2016sxe, Genolini:2016ecx}.
It remains an open problem to directly derive this anomalous transformation 
from the QFT in a new minimal supergravity background.  Returning to our present problem, 
the QFT is in any case coupled to an $\mathcal{N}=2$ conformal supergravity background, and 
for  the $\mathcal{N}=2$ topological twist we find no anomaly. 
In particular our topologically twisted supergravity theory, \emph{formally} at least, defines a topological 
theory. We discuss this further in section \ref{SecFill} and section \ref{SecDiscussion}.


\section{Geometric reformulation}\label{SecGeometry}

In this section we present a geometric reformulation of the bulk supersymmetry equations. 
 In section \ref{SecSp1Structure} we describe how (twisted) differential forms built out of bilinears in the bulk spinor define a twisted $Sp(1)$ structure on $\B_5$,
 and 
in section \ref{SecSp1DiffSys} we then derive a set of first order differential constraints on this structure. 
On the conformal boundary this restricts to the quaternionic K\"ahler structure that exists
on any oriented Riemannian four-manifold $(M_4,g)$, described in section \ref{SecTT}. We also 
discuss some general aspects of the filling problem in section \ref{SecFill}. 

\subsection{Twisted \texorpdfstring{$Sp(1)$}{Sp(1)} structure}\label{SecSp1Structure}

Recall from section \ref{SecRomans} that the bulk spinor $\epsilon$ of the Romans $\mathcal{N}=4^+$ theory is originally a quadruplet of 
spinors. These split into two doublets $\epsilon^\pm$, with eigenvalues $\pm \ii$ under $\Gamma_{45}$ (see equation (\ref{pmeigen})).
Beginning in section \ref{SecTT}, we worked in a truncated theory in which $\mathcal{B}^\pm=0$ and 
$\epsilon^-=0$. We may then define
\beq\label{zetadef}
\epsilon^+ \ = \ \begin{pmatrix} \zeta \\  -\zeta^c\end{pmatrix}~,
\eeq
where $\zeta$ is a spinor on $\B_5$, and recall that $\zeta^c\equiv\mathscr{C}\zeta^*$. 
Equation (\ref{zetadef}) is the solution to the symplectic Majorana condition 
$(\epsilon^+)^c=\epsilon^+$.  More globally, 
and as on the conformal boundary $M_4$, the spinor $\epsilon^+$ in (\ref{zetadef}) 
is a $Spin_\mathscr{G}$ spinor, where $\mathscr{G}=SU(2)_R$ -- 
see section~\ref{SecTT}.

With this notation we may define the following (local) differential forms 
\begin{equation}
\label{eq:Bilinears}
\begin{split}
S & \ \equiv \ \bar\zeta\zeta~, \qquad \qquad \quad \ \  \qquad \BK \ \equiv\  \frac{1}{S}\bar\zeta\gamma_{(1)}\zeta~, \\
\BJ^3 &\ \equiv \ \frac{\ii}{S}\bar\zeta\gamma_{(2)}\zeta, \ \quad \quad \BJ^2 + \ii \BJ^1 \ \equiv  \ \frac{1}{S}\bar\zeta^c\gamma_{(2)}\zeta \, ,
\end{split}
\end{equation}
where in our Hermitian basis of Clifford matrices recall that a bar denotes Hermitian conjugate.
There are a number of global comments to make. First, as in the discussion in section \ref{SecTT}, 
the fact that $\zeta$ is globally a twisted spinor, rather than a spinor, means that 
(\ref{eq:Bilinears}) in general only locally defines an $SU(2)\cong Sp(1)$ structure.\footnote{A general discussion of global $Sp(1)$ structures on five-manifolds may be found in 
\cite{CS}.} More globally, the $\BJ^I$ are twisted via the $SU(2)_R$ symmetry, transforming as a triplet. 
We shall call this a \emph{twisted} $Sp(1)$ structure. 
Another comment is that 
in any case the structure is well-defined only where $\zeta\neq0$. In general there may be solutions 
to the spinor equations where $\zeta=0$ on some locus. We should hence more precisely define 
$\B_5^{(0)}\equiv \B_5\setminus \{\zeta=0\}$, so that (\ref{eq:Bilinears}) is well-defined 
on $\B_5^{(0)}$. One will then need to impose certain boundary conditions on this structure,
near $\{\zeta=0\}$,  in order that the solution on $\B_5$ is appropriately regular.
The bilinears (\ref{eq:Bilinears}) define a twisted $Sp(1)$ structure on $\B_5^{(0)}$. 

The expansion of the spinor (\ref{genspinorexp}) implies that near the conformal boundary
\beq
\zeta \ = \ z^{-1/2} \chi +z^{3/2} \left(\tfrac{1}{48} R \right)\chi +z^{5/2}\left( - \tfrac{1}{24} \log z \, \diff R + \tfrac{\ii}{2}\mafive + \tfrac{1}{2}\diff \Xs + \tfrac{1}{48}\diff R\right)\cdot\chi + o(z^3)~,
\eeq
where $\chi$ is the boundary spinor defined in section \ref{SecTT}. In particular for the topological twist 
this is constant, with constant square norm $\bar\chi\chi=c^2$ (see equations (\ref{spinorexpl}), (\ref{definechi})). Without loss of generality 
we henceforth set $c=1$, so that 
\beq
S  \ =\  \frac{1}{z} + \frac{z}{24} R + o(z^{5/2})~.\label{Sexp}
\eeq
In particular notice that $\zeta\neq 0$ near to the conformal boundary at $z=0$.

\subsection{Differential system}\label{SecSp1DiffSys}

Starting from the bulk Killing spinor equations (\ref{BulkGravitino}),  (\ref{BulkDilatino}) one can derive 
a system of differential equations for the twisted $Sp(1)$ structure  (\ref{eq:Bilinears}). In the notation (\ref{zetadef})
the spinor equations read
\begin{align}
\nabla_{\mu}\zeta & \ =  \  - \tfrac{\ii}{2}\cA_{\mu}\zeta + \tfrac{\ii}{{2}} \left(\cA^1_{\mu} - \ii \cA^2_{\mu}\right)\zeta^c - \tfrac{\ii}{{2}} \cA^3_{\mu}\zeta + \tfrac{1}{3} \big( X + \tfrac{1}{2} X^{-2} \big) \gamma_\mu\zeta \nn\\
&\ \ \ + \tfrac{\ii}{24}  X^{-1} (\cF_{\nu\rho}^1-\ii \cF^2_{\nu\rho}) ( \gamma_\mu{}^{\nu\rho} - 4 \delta_\mu^\nu \gamma^\rho )\zeta^c - \tfrac{\ii}{24}  \big( X^{-1}\cF^3_{\nu\rho} + X^2 \mathcal{F}_{\nu\rho} \big) ( \gamma_\mu{}^{\nu\rho} - 4 \delta_\mu^\nu \gamma^\rho )\zeta
\, ,\nn\\[5pt]
0 &\ = \ \tfrac{{3}}{2} \ii\,  X^{-1} \partial_\mu X \gamma^\mu \zeta +{\ii}\big( X - X^{-2} \big)\zeta - \tfrac{1}{8} X^{-1} (\cF_{\mu\nu}^1-\ii \cF^2_{\mu\nu}) \gamma^{\mu\nu}\zeta^c \nn \\
& \ \ \ + \tfrac{1}{8} ( X^{-1}\cF^3_{\mu\nu} - {2} X^2 \mathcal{F}_{\mu\nu} )\gamma^{\mu\nu}\zeta\, .
\end{align}
As in section \ref{SecRomans}, it will be convenient to introduce  the real one-form
\beq
\cC \ \equiv \ \ii\cA~.
\eeq
Using these equations, a standard calculation\footnote{For example, see \cite{Alday:2015jsa}.} leads to
\begin{align}
X^{-2} \BK \ = & \ \diff \log (XS) +\cC~,\label{Kdiff}
\end{align}
together with the triplet of equations
\begin{align}
	\diff (S \BJ^I) \ = \ & - \cC\wedge S \BJ^I + (2X+X^{-2}) \BK \wedge S\BJ^I + \epsilon^{I}_{\ JK} \cA^J \wedge S \BJ^K \nn\\
	& + \tfrac{1}{2} X^{-1}S\, (*\cF^I +  \BK\wedge \cF^I) \, . \label{BilJeqns}
\end{align}
Here the Hodge dual is constructed from the volume form $\vol_5=-\mathcal{K}\wedge \vol_4$, where 
$\vol_4\equiv \frac{1}{2}\BJ^I \wedge \BJ^I$ (no sum over $I$). The sign here is chosen to match our 
earlier choice of orientation, via (\ref{volform}), as we shall see shortly.

We may read the first equation (\ref{Kdiff}) as determining the one-form $\cC$ in terms of geometric data and the function $X$:
\beq
\cC \ = \ X^{-2}\BK - \diff \log (XS)~.\label{solvecC}
\eeq
In particular, the associated flux is then
\beq
\cG \ \equiv \ \diff\cC \ = \ \ii\cF \ = \ \diff(X^{-2}\BK)~.
\eeq
Substituting (\ref{solvecC}) into (\ref{BilJeqns}), the latter simplifies to
\beq
\diff\BJ^I \ = \ \epsilon^I_{\ JK}\cA^J\wedge\BJ^K  + (\diff\log X + 2X \BK)\wedge \BJ^I + \tfrac{1}{2}X^{-1}(*\cF^I + \BK\wedge \cF^I)~.\label{dJIbulk}
\eeq

Recall that in the original Lorentzian theory $\cA$ is a $U(1)_R$ gauge field. In the real Euclidean 
section we have defined $\cC=\ii\cA$, which is a real one-form, but there is then a residual
part of the (complexified) gauge symmetry $\cC\rightarrow \cC - \diff\lambda$, where 
$\lambda$ is a global real function. The fields transform as follows:
\beq
\zeta \ \rightarrow \ \ex^{\lambda/2}\zeta~, \qquad S \ \rightarrow \ \ex^\lambda S~, \qquad 
\cC \ \rightarrow \ \cC - \diff\lambda~,
\eeq
with everything else invariant. In particular it is immediate to see that  (\ref{solvecC}), (\ref{dJIbulk}) 
are invariant under these gauge transformations. In our boundary value problem recall that 
we fixed $\cC\mid_{M_4}=0$, and in order to preserve this gauge condition on the conformal 
boundary one should restrict to gauge transformations that vanish there, so that 
$\lambda\mid_{M_4}=0$. With this caveat, one might use this gauge freedom to effectively remove 
one of the functional degrees of freedom.

Let us look at the asymptotic form of the differential conditions near the conformal boundary at $z=0$. 
Recalling the Fefferman-Graham expansion of the fields (\ref{Xexp})--(\ref{aexp}), together with the 
topological twist boundary conditions (\ref{TTwistBulk}), we have
\begin{align}
X \ = & \ 1 -\tfrac{1}{12} z^2\log z  \, R + z^2 \Xs+ o(z^2)~, \nn\\
 \qquad \cA^I \ = &  \ A^I- \tfrac{1}{4}z^2\log z \, \Jb^I_{mn}\nabla_j R^{mnj}_{\ \ \ \ \,  i}\, \diff x^i   + z^2 \aIfive+o(z^2)~,  \nn\\
 \qquad \cC \ = & \ z^2\, \ii\mafive+ o(z^2)~.
\end{align}
Here recall that $R$ is the boundary Ricci scalar, the boundary gauge field is
\beq
A^I \ = \ \tfrac{1}{2}\omega_i^{\ \overline{jk}}\Jb^I_{\overline{jk}}\, \diff x^i ~,
\eeq
where 
$\omega_i^{\ \overline{jk}}$ is the boundary spin connection, 
$R_{mnij}$ is the boundary Riemann tensor, and $\Jb^I$ are the boundary triplet of self-dual two-forms. 
The one-form $\ii\mafive$ is real.
Using also (\ref{Sexp}),  equation (\ref{Kdiff}) then implies that 
\beq
\BK   \ = \ -\frac{\diff z}{z} + z^2\big(- \tfrac{1}{12}\log z \, \diff R + \ii \mafive + \diff \Xs + \tfrac{1}{24}\diff R\big) + o(z^{5/2}) \, .\label{BKexp}
\eeq
Recall that in section \ref{SecTT} we defined the triplet of  boundary almost complex 
structures $(\Ib^I)^{i}_{\ j}\equiv \met^{ik}(\Jb^I)_{kj}$. If we define the boundary (almost) 
Ricci two-forms
\beq
\rho^I_{ij} \ \equiv \ R_{k[i}(\Ib^I)^k_{\ j]}~,
\eeq
where $R_{ij}$ is the boundary Ricci tensor, then 
similarly from the definition (\ref{JIbilinears}) we have
\begin{align}
\BJ^I \ = & \ \  \frac{1}{z^2}\Jb^I + \tfrac{1}{12}R\, \Jb^I - \tfrac{1}{2}\rho^I \nn\\
&+ z  \diff z \wedge \Ib^I \big(-\tfrac{1}{12}\log z\,  \diff R + \ii \mafive + \diff \Xs + \tfrac{1}{24}\diff R\big) + o(z^{3/2})~.\label{BJexp}
\end{align}
Here $\Ib^I(\eta)_i=(\Ib^I)^j_{\ i}\eta_j$ for a one-form $\eta$ tangent to the boundary. It is interesting to note that the  $O(1)$ terms in $\BJ^I$ above may also be written as
$\tfrac{1}{12}R\, \Jb^I - \tfrac{1}{2}\rho^I = (\mexp^2\circ J^I)$, where recall from equation (\ref{g2}) that 
$\mexp^2$ is (minus) the Schouten tensor of the conformal boundary.
From (\ref{dJIbulk}) we hence read off the leading order
the boundary equation
\beq
\diff \Jb^I \ = \ \epsilon^I_{\ JK}A^J\wedge \Jb^K~.\label{dJIboundary}
\eeq
Equation (\ref{dJIboundary}) follows from taking the skew symmetric part of (\ref{QKeqn}). In fact
since the exterior derivatives of the boundary $SU(2)$ structure $\Jb^I$ completely determine the intrinsic torsion 
(this is true for an $SU(n)$ structure in real dimension $2n$ \cite{GH}), it follows that (\ref{dJIboundary}) 
also implies (\ref{QKeqn}).

We may always choose a frame $\Sx_{\mu}^{\overline{\mu}}$ for the bulk metric on $\B_5$ such that 
\begin{align}
\BK \  & = \ -\Sx^5~, \qquad \qquad \qquad \qquad \,  \BJ^1 \ =  \ \Sx^2\wedge \Sx^3+\Sx^1\wedge \Sx^4~, \nn\\
\BJ^2 \ & =  \   \Sx^3\wedge  \Sx^1+\Sx^2\wedge  \Sx^4\, ,\qquad \,
\BJ^3 \ = \   \Sx^1\wedge  \Sx^2+\Sx^3\wedge\Sx^4~.\label{BJIframe}
\end{align}
In particular (\ref{BKexp}) identifies $\Sx^5\sim \diff z/z$ to leading order, and the sign for $\BK$ in (\ref{BJIframe}) follows
since $-\gamma_{\bar{z}}\chi = \chi$, where $\Ex^{\overline{z}}=\diff z/z$. The volume form is $\vol_5=\Sx^{12345}$.
Notice that the expansions (\ref{BKexp}), (\ref{BJexp}) imply that 
in general we may not identify 
$\mathscr{E}_\mu^{\overline{\mu}}$ near the conformal boundary with the Fefferman-Graham frame $\Ex_\mu^{\overline{\mu}}$ in (\ref{FGframe}), except to leading order.  

\subsection{Filling problem}\label{SecFill}

As explained in the introduction, given a Riemannian-four manifold $(M_4,g)$ as a fixed conformal boundary, 
at least to a zeroth order approximation in AdS/CFT
one wants to find the least action supersymmetric solution to the 
five-dimensional $\mathcal{N}=4^+$ supergravity theory, with this boundary data. 
Such a solution will be the dominant saddle point on the right hand side of (\ref{saddle}). 
In this subsection we make some comments on this problem, with further comments in 
section \ref{SecTopAdSCFT}.

As we have seen in the previous subsection, supersymmetric solutions 
on $\B_5$ are characterized geometrically in terms of a set of first order differential equations (\ref{solvecC}), (\ref{dJIbulk}) for a certain twisted $Sp(1)$ structure. In particular there is a triplet of twisted two-forms 
$\BJ^I$, $I=1,2,3$, which locally at the conformal boundary restrict to an orthonormal set of 
self-dual two-forms on $(M_4,\met)$. The differential equations become tautological on
 the boundary, and are equivalent to the fact that every oriented Riemannian four-manifold 
  has a quaternionic K\"ahler structure, {\it i.e.} has holonomy group $Sp(1)\cdot Sp(1)\cong SO(4)$. 
This differential system on $\B_5$, regarded as extending that on $(M_4,\met)$,  clearly deserves closer study. In particular, 
these are necessary conditions for a solution, but one would also like to know whether they are sufficient. 
 It should also be possible to rewrite the renormalized supergravity action (\ref{Actionfinal}) in terms of this 
geometric data. The computation in section \ref{SecVary} implies that, given any one-parameter family 
of metrics on $M_4$, the action of any family of fillings of the boundary is independent of the 
parameter. What type of invariant is this? {\it A priori} it depends on the choice of $\B_5$ filling $M_4$, 
and on the twisted $Sp(1)$ structure on $\B_5$. 

An important question is what are the global constraints on $\B_5$? As mentioned in the introduction, topologically a smooth filling $\B_5$ of $M_4$ exists if and only if the signature 
$\sigma(M_4)=0$. Moreover, as explained in section \ref{SecTopAdSCFT}, for solutions embedded in string theory 
one also needs these manifolds to be spin.\footnote{The relevant spin bordism group is $\Omega_4^{Spin}\cong \Z$, generated 
by a K3 surface, where the map to the integers is  $\sigma(M_4)/16$.} This restriction  would seem to rule out many interesting 
four-manifolds.\footnote{Although it leaves, for example, $M_4=S^1\times M_3$, for any oriented three-manifold $M_3$, 
and products of Riemann surfaces.} However, as also mentioned in the introduction, requiring $\B_5$ to be smooth is almost certainly too strong. 
Already from AdS/CFT in other contexts, it is clear that the dominant saddle point contribution can be singular, and one 
might anticipate that this is somewhat generic, at least for general $M_4$. Perhaps the appropriate question is then: 
what are the relevant singularities of $\B_5$, for a given $M_4$?\footnote{We thank S.~Gukov for discussions on this, and indeed for 
posing this precise question!} Mathematically one would need control over existence and uniqueness 
of the differential equations for the twisted $Sp(1)$ structure, for appropriate $\B_5$ (with singularities/appropriate 
internal boundary conditions) filling $M_4$. However, one might also anticipate that the supergravity action 
(\ref{Actionfinal}) could be evaluated without knowing the detailed form of the solution, but instead in terms of 
appropriate global data, and perhaps local data associated to singularities. Notice that one constraint on such 
singularities/internal boundaries is that they do not contribute to the variation of the action 
(\ref{holWardagain}) -- see the discussion after this equation.\footnote{For example, the singularities in the gravity fillings in 
\cite{Alday:2012au, Alday:2015jsa} are isolated conical singularities. Provided the radial dependence of fields near 
to the singular point are no worse than for smooth fields in flat space, such singularities will not spoil the result (\ref{holWardagain}).} 

Less ambitiously, one might also try to find explicit solutions; for example, via symmetry reduction 
so that the equations reduce to coupled ODEs. An obvious case is solutions with $\B_5=S^1\times B_4$,
where $B_4$ is a four-ball so that $\partial \B_5=M_4=S^1\times S^3$, and seek solutions invariant under 
$U(1)\times SU(2)$ (the latter acting on the left on $S^3\cong SU(2)$). 

Finally, the present problem may be contrasted to the general hyperbolic filling problem 
  described in \cite{Anderson:2004yi}. Here one also begins with an arbitrary Riemannian
  $(M_4,\met)$, which is a conformal boundary, but one instead 
  asks for the filling to be an Einstein metric of negative curvature. This 
  problem is still quite poorly understood: there are in general 
  obstructions and non-uniqueness, and one should at least impose that 
  $\met$ has a conformal representative with positive scalar curvature \cite{Witten:1999xp} (physically, so that 
  the CFT is stable).  
The geometric problem in the present paper is likely to be much better behaved: 
the equations are first order, not second order, and the solutions should be dual 
to a TQFT.  


\section{Discussion}\label{SecDiscussion}

We conclude with a discussion of ``topological AdS/CFT'' in section \ref{SecTopAdSCFT}, followed 
by various extensions and generalizations in section \ref{SecGeneralize}.

\subsection{Topological AdS/CFT}\label{SecTopAdSCFT}

 An application of the ideas developed in this paper would be to a topologically twisted version of 
 the AdS/CFT correspondence. To make quantitative comparisons between 
 calculations on the two sides, as in (\ref{saddle}) (appropriately interpreted), the construction needs embedding in string theory. 
This is straightforward: the Romans  theory is a consistent truncation of both Type IIB supergravity 
on $S^5$ \cite{Lu:1999bw}, and also of eleven-dimensional 
supergravity on $N_6$ \cite{Gauntlett:2007sm}, where $N_6$ are the geometries 
classified by Lin-Lunin-Maldacena \cite{Lin:2004nb}. This means that any solution 
to the five-dimensional Romans theory uplifts (at least locally -- see below) to a string/M-theory solution. 

In order to be concrete, let us focus on the case of $\mathcal{N}=4$ Yang-Mills theory.
Applying the Donaldson-Witten twist
leads to the half-twisted theory referred to in the introduction. 
For general gauge group $\mathscr{G}$
the path integral localizes \cite{Labastida:1997vq,  Labastida:1998sk} onto solutions to a non-Abelian \cite{Labastida:1995zj} version of the Seiberg-Witten equations, in which the spinor field is in the adjoint representation of $\mathscr{G}$. For $\mathscr{G}=SU(N)$,
 AdS/CFT 
should relate the large $N$ limit of this theory to an appropriate class of solutions to the Romans $\mathcal{N}=4^+$  
theory in five dimensions, uplifted on $S^5$ to give full solutions of Type IIB string theory. 
This is where the restriction that $M_4$ is spin enters: if $M_4$ is not spin then the background $SU(2)$ R-symmetry 
gauge field we turn on is not globally a connection on an $SU(2)$ bundle over $M_4$. 
On the other hand, the Type IIB solution is an $S^5$ fibration over the filling $\B_5$, where 
$S^5\subset \C^2\oplus \C$, and $SU(2)$ acts on $\C^2$ in the fundamental representation. 
Thus if $M_4$ is not spin, this associated bundle is not well-defined. This is also directly visible 
in the TQFT: for the half-twist of $\mathcal{N}=4$ Yang-Mills there are still spinors in the twisted theory, 
which only make sense if $M_4$ is spin. 

There is some discussion of the half-twisted $\mathcal{N}=4$ theory
for general gauge group  $\mathscr{G}$ in \cite{Lozano:1999ji}. In particular the (virtual) dimension of the the relevant non-Abelian monopole 
moduli space $\mathcal{M}$ may be computed using index theory, leading to 
\beq
\dim \mathcal{M} \ = \ -\tfrac{1}{4}\dim \mathscr{G} \cdot \left[2\chi(M_4)+3\sigma(M_4)\right]~.\label{dimM}
\eeq
Because of the associated fermion zero modes, the partition function 
of the theory vanishes unless the right hand side of (\ref{dimM}) is also zero. We have already seen precisely this condition in the holographic dual set-up, namely equation 
(\ref{topconstraint}). In the gravity context this followed from $\mathcal{A}$ being a global one-form, and then integrating the divergence of the VEV of the $U(1)_R$ current (the $U(1)_R$ anomaly) over a compact $M_4$ without boundary, as in~(\ref{intdivJR}). In fact
the two are directly related, since the virtual dimension (\ref{dimM}) of $\mathcal{M}$ computed in field theory is 
proportional to this integrated $U(1)_R$ anomaly. In the current holographic set-up, we can see this explicitly by first noting that 
for the large $N$ limit of the $\mathscr{G}=SU(N)$ half-twisted $\mathcal{N}=4$ Yang-Mills theory, a standard AdS/CFT 
formula fixes the dual effective five-dimensional Newton constant as
\beq
\frac{1}{\kappa_5^2} \ = \ \frac{N^2}{4\pi^2}~.\label{kappa5}
\eeq
This fixes the overall normalization of the supergravity action. In the large $N$ limit, using (\ref{intdivJR})  
we may then write
\beq
\dim \mathcal{M} \ = \ 2{\ii}\int_{M_4}\diff *_4\langle\, \JRcur \rangle~,\label{holdimM}
\eeq
in terms of the integrated (holographic) $U(1)_R$ anomaly. 

Another important observation is that (\ref{dimM}) is independent of the topology of the gauge bundle over $M_4$,
 unlike the corresponding case for Donaldson theory (pure $\mathcal{N}=2$ Yang-Mills 
with gauge group $\mathscr{G}$). Because of this, all choices of gauge bundle contribute 
to the partition function at the same time. The left hand side of (\ref{saddle}) then needs 
appropriately interpreting for such twists of four-dimensional  $\mathcal{N}=2$ SCFTs, 
as taken at face value it may be divergent. There is a standard way to deal with this,\footnote{We are again grateful to S.~Gukov for 
pointing this out.} namely to 
refine the partition function via the $U(1)_R$ charge. For example, this is discussed at the end of 
section 2 of \cite{Gukov:2017zao}, and in \cite{Gukov:2016gkn}. This should play an important role 
in making sense also of the right hand side of (\ref{saddle}), in addition to the comments 
on this in section \ref{SecFill}. For example, a very concrete case mentioned in 
the latter subsection is $M_4=S^1\times S^3$. Here the refined partition function is closely related to the Coulomb branch index, as explained in \cite{Dedushenko:2017tdw}. One might then try to reproduce  this from a dual supergravity solution for which $Y_5 = S^1\times B_4$, with $\partial Y_5 = S^1\times S^3$. 
More generally, for a four-manifold $S^1\times M_3$ with product metric both $\mathcal{E}$ and $\mathcal{P}$ vanish, and the holographic $U(1)_R$ current is conserved, as can be seen from \eqref{divJREP}. The associated conserved holographic R-charge might then 
 provide a natural holographic correspondent to the refinement of the partition function for the twisted four-dimensional SCFT. 
The AdS/CFT relation~(\ref{saddle}) in particular implies that the logarithm of the TQFT partition function, appropriately refined as above, 
scales as $N^2$ as $N\rightarrow\infty$, when it is non-zero.
On the other hand, when the right hand side of (\ref{dimM}) is positive, one obtains non-zero 
invariants in the TQFT by inserting appropriate $\mathcal{Q}$-exact operators into the path integral. 
We briefly discuss the dual holographic computation in section~\ref{SecGeneralize}. In particular, 
such insertions will change the boundary conditions on supergravity fields we have imposed in this paper.

As far as we are aware, computations of topological observables in the half-twisted $\mathcal{N}=4$ theory,
for general $\mathscr{G}=SU(N)$, have not been done explicitly. However, for $\GG=SU(2)$ 
the partition function and topological correlation functions have been computed 
explicitly for simply-connected spin four-manifolds of simple type \cite{Labastida:1998sk}. 
This is done by giving masses, explicitly breaking 
$\mathcal{N}=4$ to $\mathcal{N}=2$, leading to an $\mathcal{N}=2$ gauge theory 
with a massive adjoint hypermultiplet, a twisted version of the $\mc{N}=2^*$ theory. The twisted theory is still topological, 
and the relevant observables are written in terms of Seiberg-Witten invariants using the methods of \cite{Moore:1997pc}. 
Observables for the original theory are then identified with the massless limit of these formulae (when this makes sense), 
although the validity of this assertion is not completely clear. 
In any case, to compare to the holographic construction in this paper one should 
compute the large $N$ limit for gauge group $\mathscr{G}=SU(N)$. 
We note that an analogous large $N$ limit of Donaldson invariants (for pure $\mathcal{N}=2$ $SU(N)$ Yang-Mills)
has been computed in \cite{kron}. Unlike the formula (\ref{dimM}), here the dimension 
of the moduli space of instantons depends on the topology of the gauge bundle. 
One can then choose this bundle in such a way that $\dim\mathcal{M}=0$. 
The partition function is a certain signed count of the points  that make up $\mathcal{M}$, 
and the large $N$ limit was computed for a certain class 
of four-manifolds in \cite{kron}.\footnote{In particular the final 
section of \cite{kron} computes the large $N$ limit of the partition function $Z$
for a four-manifold with boundary, constructed as $S^1\times M_3$ where 
$M_3$ is a knot complement. One finds $Z\sim N \log \alpha$, where 
$\alpha$ is a certain knot invariant (the Mahler measure). }

We conclude this subsection by noting that 
similar remarks apply to twists of $\mathcal{N}=2$ SCFTs with M-theory duals. 
Indeed, an important restriction on the class of $\mathcal{N}=2$ gauge theories 
to which this holographic description applies is that they are conformal theories.\footnote{In particular this is not true of pure $\mathcal{N}=2$ Yang-Mills, from which the original Donaldson invariants are constructed.} A large number of examples arise as class $\mathcal{S}$ theories \cite{Gaiotto:2009we}, obtained by wrapping M5-branes over punctured Riemann surfaces, for which the gravity dual was found in \cite{Gaiotto:2009gz} 
using the construction of \cite{Lin:2004nb}. Romans solutions uplift on the corresponding internal spaces $N_6$ to solutions 
of M-theory \cite{Gauntlett:2007sm}.
At the level of the five-dimensional theory, all that changes is the 
formula (\ref{kappa5}) for the effective Newton constant, which in general reads \cite{Henningson:1998gx}
\beq
\frac{1}{\kappa_5^2} \ = \ \frac{a}{\pi^2}~,\label{kappaa}
\eeq
where $a$ is the $a$ central charge. In the supergravity limit recall that $a=c$. For the above-mentioned M5-brane theories the
central charge scales with $N^3$ as $N\rightarrow\infty$. Indeed, the partition function will {\it a priori} depend on both the 
choice of $\mathcal{N}=2$ SCFT that is being twisted, and also on the four-manifold $M_4$ on which it is defined. 
The choice of theory corresponds to the choice of internal space in the uplifting to ten or eleven dimensions. The structure of the dual supergravity 
solution as a fibration of the internal space over the spacetime filling of $M_4$ then implies that 
the large $N$ limits of the partition functions should also factorize. That is, the dependence on the choice of 
theory should only be visible via the central charge $a$, which via (\ref{kappaa}) fixes the overall normalization of the supergravity 
action. On the other hand, the dependence on the choice of $M_4$ is then captured by the effective five-dimensional
Romans theory we have described.\footnote{This structure can already be seen in the more general formula 
for $\dim\mathcal{M}$ given in \cite{Gukov:2017zao}. For the general class of  twisted field theories considered there, equation (2.42) of \cite{Gukov:2017zao} implies that in the large $N$ limit where $a=c$, one has $\dim\mathcal{M} = - a[2\chi(M_4)+3\sigma(M_4)]$, generalizing (\ref{dimM}).  
The central charge appears as an overall factor, at large $N$. Of course, this precisely agrees with our holographic formula (\ref{holdimM}), using 
(\ref{intdivJR}) and (\ref{kappaa}).} 

\subsection{Generalizations}\label{SecGeneralize}

We have already discussed a number of open problems and directions for future work. Here we briefly mention 
some further generalizations:
\begin{itemize}
\item Perhaps the most immediate generalization of the computations in this paper would be to the so-called 
$\Omega$-background of \cite{Nekrasov:2003rj}. Here $(M_4,g,\xi)$ is an arbitrary Riemannian four-manifold, equipped with a 
Killing vector field $\xi$. As for the pure topological twist,
this geometry also arises by coupling an $\mathcal{N}=2$ gauge theory to a certain 
background of $\mathcal{N}=2$ conformal supergravity, and is briefly mentioned at the end of section 3 of \cite{Klare:2013dka}. 
The non-zero  Killing vector $\xi$ requires turning on a boundary $B$-field: specifically one needs to take
$b^-$ (or $b^+$) proportional to the self-dual (or anti-self-dual) part of the two-form $\diff\xi^\flat$, where $\xi^\flat$ is the Killing one-form 
dual to $\xi$. Correspondingly, both boundary spinor doublets $\los^+$ and $\los^-$ are now non-zero, and one 
needs to work with the full Romans theory, rather than the truncated version with $\mathcal{B}^\pm=0$ we used from section \ref{SecTT} onwards. 
Nevertheless, the computations should not be too much more involved than those in the present paper. One expects 
the supergravity action now to depend on the choice of Killing vector $\xi$ on $M_4$, but otherwise not on the metric. 
One should thus look at metric deformations $\met_{ij}\rightarrow\met_{ij} +\delta \met_{ij}$, where 
$\mathcal{L}_\xi \, \delta \met_{ij}=0$. 
\item As mentioned in the introduction, there are three inequivalent topological twists of $\mathcal{N}=4$ Yang-Mills. 
The half-twist, relevant to this paper, was discussed in the previous subsection. The other two twists 
are the Vafa-Witten twist \cite{Vafa:1994tf}, and the twist studied by 
Kapustin-Witten in \cite{Kapustin:2006pk}. In particular in the former theory the only non-trivial observable is the partition function, 
and this has been studied for gauge group $\mathscr{G}=SU(N)$ in \cite{Labastida:1999ij}. 
These twists require the larger $SU(4)_R$ R-symmetry of the $\mathcal{N}=4$ theory, 
meaning for the holographic dual one needs to start with a Euclidean form of $\mathcal{N}=8$ gauged supergravity theory. 
Optimistically, one might hope to embed within the 
$SU(4)\sim SO(6)$ truncation of the latter theory studied in \cite{Cvetic:2000nc}, which is 
a consistent truncation of Type IIB supergravity on $S^5$, and contains  the five-dimensional Romans $\mathcal{N}=4^+$ theory (with zero $B$-field) as 
a further truncation.
\item Topological twists exist in a variety of dimensions. In three dimensions the R-symmetry group 
is $Spin(\mathcal{N})$. The analogous amount of supersymmetry to that studied in the present paper is $\mathcal{N}=4$, leading to a $ Spin(4)=SU(2)\times SU(2)$ 
R-symmetry group. On the other hand $Spin(3)=SU(2)$, and this leads to two inequivalent three-dimensional $\mathcal{N}=4$ topological twists -- see, for example, the diagram in section 1 of \cite{Geyer:2001yc}. One of these twists is closely related (by dimensional reduction on a circle) 
to the Donaldson-Witten twist. The relevant holographic construction should begin with four-dimensional $\mathcal{N}=4$ 
gauged supergravity. This contains an $Spin(4)_R$ gauge field, as required, and is a consistent truncation of eleven-dimensional supergravity 
on $S^7$ \cite{Cvetic:1999au}. The uplifted solutions should be holographically dual to twists of the ABJM theory \cite{Aharony:2008ug} 
on $N$ M2-branes, in the large $N$ limit. This is considered in~\cite{BenettiGenolini:2018iuy}.
\item Finally, in this paper we have focused exclusively on the partition function. However, in general 
TQFTs have non-trivial topological correlation functions, involving the insertion of $\mathcal{Q}$-invariant 
operators into the path integral. For example, this is true of Donaldson theory, where such insertions are required
to obtain non-zero invariants  in field theory whenever $\dim \mathcal{M}=d>0$, due to fermion zero modes.  Geometrically these  invariants arise as the integral of a $d$-form 
over $\mathcal{M}$, where this top form is itself constructed as a wedge product of certain closed forms. 
The operators are constructed via a descent procedure  \cite{Witten:1988ze}. It would be very interesting to understand the holographic dual computation
of these correlation functions. Of course, correlation functions are well studied in AdS/CFT.  In the present setting  one would again hope 
to be able to work in a truncated supergravity theory, containing the fields whose boundary values act as sources for the 
operators. Being topological, the correlation functions should be independent of the positions at which the local operators are inserted, and also independent of the metric. These statements might be proven along similar lines to the present paper. We leave this, and other interesting 
questions, for future work.
\end{itemize}


\subsection*{Acknowledgments}

\noindent 
We are very grateful to Dario Martelli and especially Sergei Gukov for detailed comments on a draft of this paper, and for discussions. 
P.~B.~G. is supported by EPSRC and a Scatcherd European Scholarship. He thanks the organisers of the Pollica Summer Workshop 2017 for hospitality, and acknowledges support from the Simons Center for Geometry and Physics, Stony Brook University during the ``Simons Summer Workshop 2017.'' He was partly supported by the ERC STG grant 306260 during the Pollica Summer Workshop. P.~R. is supported by an INFN Fellowship.


\end{document}